\documentclass[final]{dmtcs-episciences}

\usepackage{amsmath,amssymb,paralist,amsthm,eurosym}
\usepackage{placeins}
\usepackage{microtype}
\usepackage[utf8]{inputenc}
\usepackage{scalerel}
\usepackage{graphicx}

\usepackage{graphicx}
\usepackage{tikz}
\usepackage{tikz-qtree}
\usepackage{rotating}

\usetikzlibrary{arrows,matrix}
\usetikzlibrary{matrix,calc}
\usetikzlibrary{shapes.geometric}
\usetikzlibrary{positioning}
\usetikzlibrary{calligraphy}
\usetikzlibrary{fit}

\pgfdeclarelayer{background}
\pgfsetlayers{background,main}

\usetikzlibrary{decorations.pathreplacing}
\tikzset{curlybrace/.style={decoration=brace,decorate}}
\usetikzlibrary{automata}
\usetikzlibrary{calc,positioning}
\usetikzlibrary{decorations.pathmorphing}
\usetikzlibrary{shapes}
\tikzset{trinode/.style={draw,triangle,minimum width=2.0cm}}
\tikzset{snake/.style={decorate,decoration=snake}}
\tikzset{curlybrace/.style={decoration=brace,decorate}}
\tikzset{triangle/.style={regular polygon,regular polygon sides=3}}
\tikzset{edge from parent path={(\tikzparentnode) -- (\tikzchildnode.north)}}
\usetikzlibrary{arrows.meta}

\usepackage{wrapfig}
\usepackage[rightcaption]{sidecap}
\usepackage[colorinlistoftodos]{todonotes}
\usepackage{mathtools}
\usepackage{multicol} 
\usepackage{bbold}
\DeclareMathAlphabet{\mathbbold}{U}{bbold}{m}{n}
\DeclareSymbolFontAlphabet{\mathbb}{AMSb}
\usepackage{todonotes}
\usepackage{imakeidx}         

\newtheorem{theorem}{Theorem}[section]
\newtheorem{lemma}[theorem]{Lemma}

\newtheorem{corollary}[theorem]{Corollary}
\newtheorem{observation}[theorem]{Observation}
\newtheorem{example}[theorem]{Example}
\newtheorem{definition}[theorem]{Definition}

\newcommand{\cA}{\mathcal A}
\newcommand{\cB}{\mathcal B}

\newcommand{\T}{\mathrm{T}}

\newcommand{\R}{\mathrm{R}}

\newcommand{\im}{\mathrm{im}}

\newcommand{\supp}{\mathrm{supp}}

\newcommand{\wt}{\mathrm{wt}}

\newcommand{\B}{\mathsf{B}}

\newcommand{\Boole}{\mathsf{Boole}}

\newcommand{\sfL}{\mathsf{L}}

\newcommand{\Nat}{\mathsf{Nat}}

\newcommand{\0}{\mathbbold{0}}
\newcommand{\1}{\mathbbold{1}}

\newcommand{\h}{\mathrm{h}}

\newcommand{\pos}{\mathrm{pos}}
\newcommand{\prefix}{\mathrm{prefix}}

\newcommand{\rk}{\mathrm{rk}}

\newcommand{\sem}[1]{[\![#1]\!]}

\newcommand{\runsem}[1]{[\![#1]\!]^{\mathrm{run}}}
\newcommand{\initialsem}[1]{[\![#1]\!]^{\mathrm{init}}}

\newcommand{\nf}{\mathrm{nf}}

\newcommand{\tree}{\mathrm{tree}}

\usepackage[most]{tcolorbox}
\newtheorem{theorem-rect}[theorem]{Theorem}
\newtheorem{corollary-rect}[theorem]{Corollary}
\newtheorem{lemma-rect}[theorem]{Lemma}
\newtheorem{construction-rect}[theorem]{Construction}
\newtheorem{definition-rect}[theorem]{Definition}

\tcolorboxenvironment{theorem-rect}{
		arc=3pt,				
		colback=white,			
		enlarge top by=2mm,		
		enlarge bottom by=-1mm,	
		grow to left by=0mm,	
		grow to right by=0mm,	
		boxrule=0.1mm,			
		left=1.7mm,				
		right=1.7mm				
		}						

\tcolorboxenvironment{corollary-rect}{
		arc=3pt,
		colback=white,
		enlarge top by=2mm,
		enlarge bottom by=-1mm,
		grow to left by=0mm,
		grow to right by=0mm,
		boxrule=0.1mm,
		left=1.7mm,
		right=1.7mm
              }

\tcolorboxenvironment{lemma-rect}{
		arc=3pt,
		colback=white,
		enlarge top by=2mm,
		enlarge bottom by=-1mm,
		grow to left by=0mm,
		grow to right by=0mm,
		boxrule=0.1mm,
		left=1.7mm,
		right=1.7mm
              }

\tcolorboxenvironment{construction-rect}{
		arc=3pt,
		colback=white,
		enlarge top by=2mm,
		enlarge bottom by=-1mm,
		grow to left by=0mm,
		grow to right by=0mm,
		boxrule=0.1mm,
		left=1.7mm,
		right=1.7mm
              }

\tcolorboxenvironment{definition-rect}{
		arc=3pt,
		colback=white,
		enlarge top by=2mm,
		enlarge bottom by=-1mm,
		grow to left by=0mm,
		grow to right by=0mm,
		boxrule=0.1mm,
		left=1.7mm,
		right=1.7mm
              }

\pgfdeclaredecoration{dashsoliddouble}{initial}{
  \state{initial}[width=\pgfdecoratedinputsegmentlength]{
    \pgfmathsetlengthmacro\lw{.7pt+.5\pgflinewidth}
    \begin{pgfscope}
      \pgfpathmoveto{\pgfpoint{0pt}{\lw}}%
      \pgfpathlineto{\pgfpoint{\pgfdecoratedinputsegmentlength}{\lw}}%
      \pgfmathtruncatemacro\dashnum{%
        round((\pgfdecoratedinputsegmentlength-3pt)/6pt)
      }
      \pgfmathsetmacro\dashscale{%
        \pgfdecoratedinputsegmentlength/(\dashnum*6pt + 3pt)
      }
      \pgfmathsetlengthmacro\dashunit{3pt*\dashscale}
      \pgfsetdash{{\dashunit}{\dashunit}}{0pt}
      \pgfusepath{stroke}
      \pgfsetdash{}{0pt}
      \pgfpathmoveto{\pgfpoint{0pt}{-\lw}}%
      \pgfpathlineto{\pgfpoint{\pgfdecoratedinputsegmentlength}{-\lw}}%
      \pgfusepath{stroke}
    \end{pgfscope}
  }
}

\tikzset{small circle/.style={circle, draw=black, inner sep=0pt,outer sep=0pt, minimum size=2.5pt}}

\title[Run supports and initial algebra supports]{Run supports and initial algebra supports of weighted automata}
\author[Manfred Droste and Heiko Vogler]{Manfred Droste\affiliationmark{1} \and Heiko Vogler\affiliationmark{2}}
\affiliation{Leipzig University, Leipzig, Germany\\
Technische Universit\"at Dresden, Dresden, Germany}

\begin{document}
\publicationdata{vol. 28:3}{2026}{15}{10.46298/dmtcs.14265}{2024-09-16; 2024-09-16; 2026-08-25}{2026-09-22}
\received{2024-09-16}
\revised{2026-08-25}
\maketitle
\begin{abstract}
\vspace{0.2cm}
We consider weighted automata over words and over trees where the weight algebras are strong bimonoids, i.e., semirings which may lack distributivity. It is well known that, for each such weighted automaton, its run semantics and its initial algebra semantics can be different, due to the absence of distributivity. Here we investigate the question under which conditions on a zero-sum-free strong bimonoid the support of the run semantics equals the support of the initial algebra semantics. We prove a characterization of this equality both for weighted automata over words and for weighted automata over trees in terms of two weakened distributivity laws for the strong bimonoids which are required to hold only for expressions evaluating to zero. This provides a natural extension of two classical results on the coincidence of the run semantics and the initial algebra semantics. We also consider shortly the images of the two semantics functions.
\end{abstract}

  \section{Introduction}

  Weighted automata \cite{sch61} are a natural generalization of finite-state automata.
A weighted automaton assigns to each of its transitions a weight, and its semantics generalizes the Boolean-valued recognition of an input structure to a computation of a weight for that input structure in some weight algebra. In this way, a weighted automaton computes, for each input structure, a weight in the weight algebra.

Weighted automata have been investigated for different kinds of input structures, like words \cite{sch61}, infinite words \cite{drokus06}, trees \cite{inafuk75}, pictures \cite{bozgra05}, traces \cite{drogas99}, nested words \cite{mat08,mat10}, timed words \cite{qua09} and for different classes of weight algebras like rings \cite{sch61}, fields \cite{berreu82}, semirings \cite{eil74}, lattices \cite{inafuk75}, multioperator monoids \cite{kui98}, strong bimonoids \cite{cirdroignvog10,drostuvog10,rad10}, valuation monoids \cite{dromei12}, and hemirings \cite{drokui13}. 
Here we focus on weighted automata over words \cite{eil74,salsoi78,wec78,kuisal86,berreu88,sak09,drokuivog09,drokus21} and weighted automata over trees \cite{inafuk75,berreu82,kui98,esikui03,fulvog09new,fulvog26}.
As weight algebras we consider strong bimonoids, which can be understood as semirings which need not obey the distributivity laws. 

Central in automata theory are the languages accepted by the automata. For weighted automata, these languages are called \emph{supports} and are given by the input structures for which the weighted automaton computes a non-zero value. It is the goal of this paper to compare two standard computation methods when for given input structures (words or trees) they both compute a non-zero value. In particular, we obtain a characterization result for 
zero-sum-free
strong bimonoids when for this question the most standard computation method, which typically has an exponential complexity, can be replaced by the second method, needing in general only a number of operations linear in the size of the input.

Strong bimonoids which are not semirings abound. They often occur in multi-valued logics, e.g., in Gödel logics, {\L}ukasiewicz-logics, Post-logics, cf. \cite{got01}. For instance, bounded lattices are strong bimonoids (and in lattice theory distributivity is considered a strong assumption, cf. \cite{gra03}). Quantum automata and quantum logic with values in orthomodular lattices, 
which are often not distributive, were investigated, e.g., in \cite{Yin00, Qiu04, Li10}.
The importance of non-distributive lattices for quantitative model checking has been stressed in \cite{mall05}. 
For surveys on lattice-valued automata, see \cite{rah09, fulvog26}.

Strong bimonoids which are not semirings also arise via functions on monoids, cf. \cite{bla53}, including nonlinear operators in functional analysis and mathematical physics, cf. \cite[p.3842]{sch97}.
Naturally, standard polynomials with usual addition and composition form well-known examples of right-distributive strong bimonoids which are not semirings, cf. \cite{pil77}.
Weighted automata over strong bimonoids were investigated for words in, e.g., \cite{drostuvog10,cirdroignvog10,drovog12} and for trees in, e.g., \cite{rad10,droheu15, fulkosvog19,fulvog26}. 

For a weighted automaton $\cA$ we can distinguish (at least) two computation methods: the run semantics and the initial algebra semantics. In the run semantics (\cite[VI.6]{eil74} for the case of words, and \cite[p.~83]{fulvog26} for the case of trees), to each position of an input structure an arbitrary  state of $\cA$ is assigned. Taking the assignment to the direct predecessors into account, this assignment (called \emph{run}) determines for each position a transition. The weight of a run is computed by multiplying inital weights, the weights of these transitions and final weights. Finally, the weights of all the runs on the input structure are summed up. Summation $\oplus$ and multiplication $\otimes$ are provided by the underlying strong bimonoid $\B=(B,\oplus,\otimes,\0,\1)$. In this way, the run semantics is a mapping $\runsem{\cA}$ of type $\runsem{\cA}: \Gamma^* \to B$ (for the word case; $\Gamma^*$ is the set of words over an alphabet $\Gamma$) and $\runsem{\cA}: \T_\Sigma \to B$ (for the tree case; $\T_\Sigma$ is the set of trees over a ranked alphabet $\Sigma$). 

The notion of initial algebra semantics is due to the seminal paper \cite{gogthawagwri77}.
In the initial algebra semantics of a weighted automaton  $\cA$ with state set  $Q$, one uses state vectors  $u \in B^Q$ to describe
the dynamically evolving weighted behavior of  $\cA$. For the word case, this evolvement starts with the state vector of initial weights. Each input symbol $a \in \Gamma$ respectively $k$-ary $\sigma \in \Sigma$ (including $k=0$) is interpreted as a unary respectively $k$-ary operation on $B^Q$; in the tree case, the evolvement starts at the nullary operations. Intuitively, the $q$-component of the application of such an operation is the sum of the weights of all transitions on $a$ and $\sigma$, respectively, which end up in $q$.
After executing all symbols from the given word respectively tree, the weight vector obtained is multiplied with the final weight vector of $\cA$. This results in the initial algebra semantics $\initialsem{\cA}: \Gamma^* \to B$ respectively $\initialsem{\cA}: \T_\Sigma \to B$. It can be shown that this yields an initial algebra semantics precisely as developed in \cite{gogthawagwri77} and thereby enables the application of fruitful algebraic methods for this setting. Moreover, the initial algebra semantics has the advantage that, given a strong bimonoid  $\B$ with computable operations, the value $\initialsem{\cA}(w)$ for $w \in \Gamma^*$ respectively $\initialsem{\cA}(\xi)$ for
$\xi \in \T_\Sigma$ can be computed with a number of additions or multiplications which is linear in the length of $w$ respectively the size of $\xi$ (in contrast to the algorithm following the definition of the run semantics), cf. \cite[Thm.~6.1.1]{fulvog26}. 

For a weighted automaton over a strong bimonoid  $\B$, the two semantics can be different,  for the word case cf. \cite[Ex.~25,26]{drostuvog10} and \cite[Ex.~3.1]{cirdroignvog10}, and for the tree case cf., e.g., \cite[Sec.~6.2]{fulvog26}. This difference is due to the absence of distributivity and shows a remarkable difference between words and trees. It can be precisely described as follows.
We have $\runsem{\cA}= \initialsem{\cA}$  for each weighted automaton $\cA$  over  $\B$ if and only if, in the word case, the strong bimonoid  $\B$ is right-distributive \cite[Lm.~5]{drostuvog10}, 
and in the branching tree case, $\B$  is distributive  \cite[Lm.~4.1.13]{bor04b} and  \cite[Thm.~4.1]{rad10}.

In this paper, we deal with the supports of the two semantics of weighted automata and, briefly, also with the images of these functions.  
The support of the run semantics of a weighted automaton $\cA$ is the set $\supp(\runsem{\cA})$ of all input structures which are not mapped to the neutral element $\0$ of $\B$; similarly the support of the initial algebra semantics is defined.
Supports of the semantics of  weighted automata have been investigated since the beginnings of weighted automata theory, due to their connections to formal language theory. It is known, for instance, that the run support of a weighted automaton might be not recognizable and even not context-free, cf. \cite[chapter III]{berreu88}.
On the other hand, for each weighted automaton over a zero-sum free commutative semiring, the run support is recognizable \cite[Thm.~3.1]{kir11}. This even holds if the assumption on distributivity is dropped \cite{droheu15,goe17,fulhervog18}.

We will prove a characterization of the property:
\begin{equation}\label{eq:equality-of-supports}
  \text{for each weighted automaton $\cA$ over $\B$ we have } \ \supp(\runsem{\cA}) = \supp(\initialsem{\cA}) \tag{$\mathrm{Supp}$}
\end{equation}
in terms of properties of the underlying strong bimonoid $\B$, assuming that  $\B$  is zero-sum-free. Up to now, only sufficient conditions for Property~\eqref{eq:equality-of-supports} are known, which we recall here and in theorems below (the references to the literature are discussed there):
\begin{compactitem}
\item for the word case: $\B$ is positive or right-distributive  (cf. Theorem~\ref{thm:sufficient-cond-wsa-equality-supp})
\item for the tree case:  $\B$ is positive or distributive  (cf. Theorem~\ref{thm:sufficient-cond-wta-equality-support}).
\end{compactitem}

The proofs in these cases are rather straightforward, which shows that the assumptions of positivity or distributivity are very strong. We wish to weaken these assumptions.
We will show that restricting the distributivity law to expressions evaluating to  $\0$, together with the usual zero-sum-freeness of addition, will serve our purposes. 
In particular, we use the concept of zero-right-distributivity, already shortly introduced in \cite{ghovog23}, and we introduce the new concept of zero-distributivity.
Then we obtain characterization results for zero-sum-free strong bimonoids as follows:

{\bf Theorem \ref{thm:wsa-equivalence}:}  Let $\Gamma$ be an alphabet and $\B$ be a zero-sum-free strong bimonoid. Then the following statements are equivalent.
  \begin{compactenum}
  \item[(1)] $\B$ is zero-right-distributive.
  \item[(2)] For each weighted automaton $\cA$ over  $\Gamma$  and  $\B$, we have $\supp(\runsem{\cA}) = \supp(\initialsem{\cA})$.
  \end{compactenum}

{\bf Theorem  \ref{thm:bi-strongly-zsf-equiv-equ-supp-b}:} Let  $\Sigma$  be a branching ranked alphabet and  $\B$  be a zero-sum-free strong bimonoid.
    Then the following two statements are equivalent.
    \begin{compactenum}
    \item[(1)]  $\B$ is zero-distributive.
     \item[(2)] For each weighted tree automaton $\cA$ over $\Sigma$ and $\B$, we have $\supp(\initialsem{\cA}) = \supp(\runsem{\cA})$.
     \end{compactenum}

As mentioned above, the equality of the two semantics for any weighted automaton (over words resp. over trees as inputs) has been characterized by the strong bimonoid  $\B$  being right-distributive resp. distributive. In both conditions (2) above, we weaken the equality of the semantics
to equality of the supports, i.e., intuitively, equality of the semantics values up to being zero or non-zero.
Our results show that for zero-sum-free strong bimonoids $\B$, this corresponds precisely to weakening (right-) distributivity of $\B$  to (right-) distributivity at zero.

The following implications hold for each zero-sum-free strong bimonoid $\B$:
\begin{center}
$\B$ positive \ $\Rightarrow$ \ $\B$ zero-distributive \ $\Rightarrow$ \ $\B$ zero-right-distributive.
\end{center}
In Section \ref{sec:zsf-strong-bimonoids} we show natural examples for each of these classes, and we show that each of these implications is strict. 
In particular, natural classes of bounded lattices provide zero-distributive strong bimonoids which are neither positive nor right-distributive; hence to these lattices the above results apply (but not the known results for semirings or for positive or right-distributive strong bimonoids).  
   
Our proofs for Theorems \ref{thm:wsa-equivalence} and \ref{thm:bi-strongly-zsf-equiv-equ-supp-b} 
proceed through a careful comparison of computations of the run semantics and the initial algebra semantics of a weighted automaton  $\cA$; in particular for trees, this also involves an analysis of the structure of the trees involved. 
The new concepts of zero-(right-)distributivity for zero-sum-free strong bimonoids allow us, in a product,
     \begin{compactitem}
     \item to extend a factor by a sum with the factor as one of its summands and
       \item to reduce a sum occurring as a factor in the product into one of its summands
     \end{compactitem}
     without changing the property of the product of being $\not=\0$.
Note that for trees, in comparison to words, we need a stronger assumption on the underlying strong bimonoid $\B$ in order to achieve the equality condition for the run and initial algebra supports for all weighted tree automata $\cA$. 

As a consequence, we obtain for zero-right-distributive and zero-sum-free computable strong bimonoids and any weighted automaton ${\cA}$  that it can be decided for any given word $w$ in linear time (in the length of $w$) whether its run semantics assigned by  ${\cA}$  is non-zero; the same complexity holds for zero-distributive and zero-sum-free computable strong bimonoids and weighted tree automata 
(cf. Corollary~\ref{cor:comm-zsf-wsa=wta-b}).

We note that lattice-valued automata have been investigated in the area of model-checking, cf., e.g.,
\cite{mall05}; hence the above computability results could be applied, for natural classes of bounded lattices, to these automata.

Finally, we show that the characterization theorems mentioned above of the equality of the two semantics for weighted automata can be strengthened  as follows: if for all weighted automata  $\cA$ the images of the two functions  $\runsem{\cA}$ and $\initialsem{\cA}$ coincide, then for all weighted automata  $\cA$, the two functions  $\runsem{\cA}$ and $\initialsem{\cA}$ coincide; see Theorems \ref{thm:wsa-im-equ-implies-equ} and \ref{thm:wta-im-equ-implies-equ}. This, at first sight surprizing, result follows easily as known examples can be exploited to show that the assumption on the images implies that the strong bimonoid  $\B$ is right-distributive in the word case respectively distributive in the tree case.

\section{Preliminaries}
  \paragraph{Notations.}
We let $\mathbb{N} = \{0,1,2,\ldots\}$ and $\mathbb{N}_+ = \mathbb{N}\setminus \{0\}$. For every $i,j \in \mathbb{N}$, we let $[i,j]$ denote the set $\{\ell \in \mathbb{N} \mid i \le \ell \le j\}$. We abbreviate $[1,j]$ by $[j]$.

For each set $A$, the cardinality and the $n$-fold Cartesian product ($n \in \mathbb{N}$) are denoted by $|A|$ and $A^n$, respectively.

Let $A$ and $B$ be sets. The set of all mappings from $A$ to $B$ is denoted by~$B^A$.

\paragraph{Words.}
An \emph{alphabet} $\Gamma$ is a finite non-empty set. A \emph{word (over $\Gamma$)}, often also called a \emph{string}, is a finite sequence $w = a_1 \cdots a_n$ where $n \in \mathbb{N}$ and $a_i \in \Gamma$ ($i \in [n]$); $n$ is the length of $w$. The empty word, denoted by $\varepsilon$, has length $0$. The set of \emph{prefixes of $w$} is the set $\prefix(w) = \{\varepsilon\} \cup \{a_1 \cdots a_i \mid i \in [n]\}$. We let $\Gamma^*$ denote the set of words over $\Gamma$.

  \paragraph{Well-founded induction.}
Here we recall terminating relations and well-founded induction (cf. \cite[Sect.~2.2]{baanip98}, which we will use for the analysis of cut sets in trees.

Let $A$ be a set and $\succ \, \subseteq A \times A$. We write $a \succ b$ instead of $(a,b) \in \, \succ$. As usual, $\succ^*$  denotes the reflexive, transitive closure of $\succ$.  An element $a \in A$ is said to be \emph{in normal form (with respect to $\succ$)} if there does not exists a $b \in A$ such that $a \succ b$.  For every  $a,b \in A$, $b$ is a \emph{normal form of $a$} if $a \succ^* b$ and $b$ is in normal form.  We denote the set of all normal forms of $a$ by $\nf_\succ(a)$. For each $A' \subseteq A$ we put $\nf_\succ(A') = \{\nf_\succ(a) \mid a \in A'\}$.

We say that $\succ$ is \emph{terminating} if there does not exist a family $(a_i \mid i\in \mathbb{N})$ of elements in $A$ with $a_i \succ a_{i+1}$ for each $i \in \mathbb{N}$.  

Let $\succ$ be terminating.  Moreover, let $P \subseteq A$ be a subset, called \emph{property}. We will abbreviate the fact that $a \in P$ by $P(a)$ and say that \emph{$a$ has the property $P$}. Then the following holds:
\begin{equation}
\Big( (\forall a \in A): \ [(\forall b \in A): (a \succ b) \rightarrow P(b)] \rightarrow P(a)\Big) \rightarrow \Big((\forall a\in A): P(a)\Big)\label{equ:well-founded-induction} \enspace,
\end{equation}
where $\rightarrow$ denotes the logical implication.
The formula \eqref{equ:well-founded-induction} is called the principle of \emph{proof by well-founded induction on $(A,\succ)$}. The proof of $P(a)$ for each $a \in \nf_\succ(A)$ is the induction base and the proof of
\begin{equation}\notag
(\forall a \in A\setminus \nf_\succ(A)): \ [(\forall b \in A): (a \succ b) \rightarrow P(b)] \rightarrow P(a)
\end{equation}
is the induction step. The subformula $(\forall b \in A): (a \succ b) \rightarrow P(b)$ is the induction hypothesis. 
For instance, the classical induction on natural numbers is the proof by induction on $(\mathbb{N},\succ_\mathbb{N})$ where 
\[\succ_\mathbb{N} = \{(n+1,n) \mid n \in \mathbb{N}\}\enspace.\]
Clearly, $\nf_{\succ_\mathbb{N}}(n) =\{0\}$  for each  $n \in \mathbb{N}$.

\paragraph{Ranked alphabets and trees.}
\label{sec:ra-trees}
A {\em ranked alphabet} is a pair $(\Sigma,\rk)$, where
\begin{compactitem}
\item $\Sigma$ is an alphabet, i.e., a non-empty and finite set and
\item $\rk: \Sigma \rightarrow \mathbb{N}$ is a mapping, called \emph{rank  mapping}, such that $\rk^{-1}(0)\not=\emptyset$.
\end{compactitem}

For each $k \in \mathbb{N}$, we denote the set $\rk^{-1}(k)$ by $\Sigma^{(k)}$. Sometimes we write $\sigma^{(k)}$ to indicate that $\sigma\in\Sigma^{(k)}$.  Whenever the rank mapping is clear from the context or it is irrelevant, then we abbreviate the ranked alphabet $(\Sigma,\rk)$ by $\Sigma$.

Let $\Sigma$ be a ranked alphabet. It is 
\begin{compactitem}
\item \emph{trivial} if $\Sigma=\Sigma^{(0)}$,
  \item \emph{monadic} if $\Sigma=\Sigma^{(1)} \cup \Sigma^{(0)}$,
  \item a \emph{string ranked alphabet} if $\Sigma$ is monadic, $\Sigma^{(1)} \not= \emptyset$, and $|\Sigma^{(0)}|=1$,
    and
    \item \emph{branching} if there exists $k \in \mathbb{N}$ with $k \ge 2$ such that $\Sigma^{(k)} \not= \emptyset$.
  \end{compactitem}
  In particular,   
  a non-trivial ranked alphabet is either monadic or branching. Each string ranked alphabet is non-trivial and monadic. The following table shows some examples of ranked alphabets and their properties.

  \

  {\small
  \begin{tabular}{l|c|c|c|c}
    & $\{\alpha^{(0)}, \beta^{(0)}\}$ &  $\{\alpha^{(0)}, \gamma^{(1)}, \delta^{(1)}\}$ &  $\{\alpha^{(0)}, \beta^{(0)},\gamma^{(1)}, \delta^{(1)}\}$ & $\{\alpha^{(0)}, \sigma^{(2)}\}$\\\hline
    trivial & yes & no & no & no \\
    
    string ranked alphabet & no & yes & no & no\\
    non-trivial and monadic & no & yes &yes & no \\
                                            branching & no & no & no &yes
    \end{tabular}
  }
  
    \
    
  \begin{quote} \emph{In the rest of this paper, $\Sigma$ will denote an arbitrary but fixed ranked alphabet, if not stated otherwise.}
      \end{quote}

  Let $\Xi$ denote the set which contains three symbols: comma, opening parenthesis, and closing parenthesis. The set of \emph{trees over $\Sigma$}, denoted by $\T_\Sigma$, is the smallest set $T \subseteq (\Sigma \cup \Xi)^*$ such that (i) $\Sigma^{(0)} \subseteq T$ and (ii) for every $k \in \mathbb{N}_+$, $\sigma \in \Sigma^{(k)}$, and $\xi_1,\ldots,\xi_k \in T$, we have that $\sigma(\xi_1,\ldots,\xi_k) \in T$. Subsequently, we apply the usual structural induction on  $\T_\Sigma$  for various definitions.
  
  Let $\xi \in \T_\Sigma$. We define the \emph{set of positions of $\xi$}, denoted by $\pos(\xi)$, by induction on $\T_\Sigma$ as follows: (i)~For each $\xi \in \Sigma^{(0)}$, we let $\pos(\xi) =\{\varepsilon\}$ and (ii) for every $k \in \mathbb{N}_+$, $\sigma \in \Sigma^{(k)}$, and $\xi_1,\ldots,\xi_k \in \T_\Sigma$, we let $\pos(\sigma(\xi_1,\ldots,\xi_k)) = \{\varepsilon\} \cup \bigcup_{i \in [k]} \{iw \mid w \in \pos(\xi_i)\}$. In particular, $\pos(\xi) \subseteq (\mathbb{N}_+)^*$.
The \emph{root of $\xi$} is the position $\varepsilon$; and a position $w \in \pos(\xi)$ is a \emph{leaf of $\xi$} if there does not exist $i \in \mathbb{N}_+$ such that $wi \in \pos(\xi)$.

The \emph{lexicographic order on $\pos(\xi)$}, denoted by $\le_{\mathrm{lex}}$, and the \emph{depth-first post-order on $\pos(\xi)$}, denoted by $\le_{\mathrm{po}}$, are defined for every $w,v \in \pos(\xi)$ by
\begin{align*}
w \le_{\mathrm{lex}} v \ &\text{ iff } (w\in \prefix(v)) \vee (w <_{\mathrm{left-of}} v)\\
  w \le_{\mathrm{po}} v \ &\text{ iff } (v\in \prefix(w)) \vee (w <_{\mathrm{left-of}} v) \ \text{ where }\\
  w <_{\mathrm{left-of}} v \ &\text{ iff } (\exists u \in \prefix(w) \cap \prefix(v))(\exists i,j \in \mathbb{N}_+):\\
  &\qquad (ui \in \prefix(w)) \wedge (uj \in \prefix(v)) \wedge (i < j)
\end{align*}
We let $w <_{\mathrm{lex}} v$ if $(w \le_{\mathrm{lex}} v) \wedge (w\ne v)$, and we let $w <_{\mathrm{po}} v$ if $(w \le_{\mathrm{po}} v) \wedge (w\ne v)$. For instance, for $\xi = \sigma(\delta(\alpha,\beta),\alpha)$ with binary $\sigma$ and $\delta$ and nullary $\alpha$ and $\beta$, we have
\begin{align*}
  \varepsilon <_{\mathrm{lex}} 1 <_{\mathrm{lex}} 11 <_{\mathrm{lex}} 12 <_{\mathrm{lex}} 2 \ \ \text{ and } \ \
  11 <_{\mathrm{po}} 12 <_{\mathrm{po}} 1 <_{\mathrm{po}} 2 <_{\mathrm{po}} \varepsilon \enspace.
  \end{align*}

By induction on $\T_\Sigma$, we define the \emph{label of a tree $\xi \in \T_\Sigma$ at a position $w \in \pos(\xi)$}, denoted by $\xi(w)$, and the \emph{subtree of $\xi \in \T_\Sigma$ at $w\in \pos(\xi)$}, denoted by $\xi|_w$, as follows. (i) If $\xi \in \Sigma^{(0)}$, then $\xi(\varepsilon) = \xi$ and $\xi|_\varepsilon=\xi$, and (ii) if $\xi= \sigma(\xi_1,\ldots,\xi_k)$ with $k \in \mathbb{N}_+$, then $\xi(\varepsilon) = \sigma$ and $\xi|_\varepsilon=\xi$, and for every $i \in [k]$  and  $v \in \pos(\xi)$, we let
$\xi(iv) = \xi_i(v)$ and $\xi|_{iv} = \xi_i|_v$.
    
For each subset $\Delta \subseteq \Sigma$, we abbreviate the set $\{w \in \pos(\xi) \mid \xi(w) \in \Delta\}$ by $\pos_\Delta(\xi)$. Thus, $\pos_{\Sigma^{(0)}}(\xi)$ is the set of leaves of $\xi$.

\paragraph{Strong bimonoids.}
A structure  $\B =(B,\oplus,\otimes,\0,\1)$  is said to be a \emph{strong bimonoid}, if  $(B,\oplus,\0)$  is a commutative monoid,  $(B,\otimes,\1)$ is a monoid, and $\0 \otimes b = b \otimes \0 = \0$ for each  $b \in B$.
The strong bimonoid  $\B$  is \emph{right-distributive}, if $(a \oplus b) \otimes c = a \otimes c \oplus b \otimes c$ for all  $a,b,c \in B$. Left-distributivity is defined analogously. A \emph{semiring} is a strong bimonoid which is left- and right-distributive. The strong bimonoid is \emph{commutative}, if the multiplication operation is commutative.

Next we give a few examples of commutative strong bimonoids some of which are not semirings. For various further examples, we refer to reader to \cite{drostuvog10, drovog12, fulvog26}.

\begin{example}\rm \label{ex:strbim} The following are examples of commutative strong bimonoids:
  \begin{enumerate}
  \item The Boolean semiring $\Boole = (\mathbb{B},\vee,\wedge,0,1)$ where $\mathbb{B}= \{0,1\}$ (the truth values) and $\vee$ and $\wedge$ denote disjunction and conjunction, respectively.
    
  \item The strong bimonoid $\Nat_{+,\min} = (\mathbb{N}_\infty,+,\min,0,\infty)$ where $\mathbb{N}_\infty=\mathbb{N} \cup \{\infty\}$ and $+$ and $\min$ are the usual addition and minimum on natural numbers, respectively, extended in the natural way to $\mathbb{N}_\infty$.
    
  \item The plus-plus strong bimonoid of natural numbers \cite[Ex.~2.3]{drofulkosvog21} $\Nat_{+,+} = (\mathbb{N}_\0,\oplus,+,\0,0)$ where  $\mathbb{N}_\0 = \mathbb{N} \cup \{ \0 \}$ for some new element $\0 \not\in \mathbb{N}$. The binary operation $\oplus$, if restricted to $\mathbb{N}$, and the binary operation $+$, if restricted to  $\mathbb{N}$,  are the usual addition on natural numbers (e.g. $3 + 2 = 5$). Moreover, $\0 \oplus x = x \oplus \0 = x$  and  $\0 + x = x + \0 = \0$  for each $x \in \mathbb{N}_\0$.
  \item Any bounded lattice  $\sfL = (L,\vee,\wedge,\0, \1)$ (so $\0 \leq x \leq \1$ for each $x \in L$) where $\vee$ and $\wedge$ are the join and meet, respectively. This strong bimonoid is a semiring if and only $\sfL$ is a distributive lattice.
  \end{enumerate}
    \end{example}

\paragraph{Weighted sets.}
Let $\B = (B,\oplus,\otimes,\0,\1)$ be strong bimonoid and $A$ a set. A mapping $r: A \to B$ is called \emph{($B$-)weighted set}. In case that $A = \Gamma^*$ for some alphabet $\Gamma$ or $A=\T_\Sigma$ for some ranked alphabet $\Sigma$, we call $r$ a \emph{weighted language} and \emph{weighted tree language}, respectively.

Let $r: A \to B$ be a weighted set. The \emph{support of $r$}, denoted by $\supp(r)$, is the set $\{a \in A \mid r(a) \not= \0\}$.

\section{Zero-sum-free strong bimonoids}
\label{sec:zsf-strong-bimonoids}

In this section we will define our new notion of zero-distributive strong bimonoids, and we compare them with positive strong bimonoids. 

Let  $\B =(B,\oplus,\otimes,\0,\1)$  be a strong bimonoid. 
We say that  $B$  is
\begin{compactitem}
\item \emph{zero-sum-free}, if for all  $a, b \in B$  we have:
  $a \oplus b = \0$  iff  $a = b = \0$.
 \item \emph{zero-divisor-free}, if for all  $a, b \in B$  we have:
    $a \otimes b = \0$  iff  $a = \0$ or $b = \0$.
    \item \emph{positive}, if it is zero-sum-free and zero-divisor-free.
  \item \emph{zero-right-distributive} (cf. \cite[Lm.~4.3]{ghovog23}), if  for all  $a, b, c \in B$  we have:
  $(a \oplus b) \otimes c = \0$  iff  $a \otimes c \oplus b \otimes c = \0$.
  \item \emph{zero-distributive}, if  for all  $a, b, b', c \in B$  we have:
 $a \otimes (b \oplus b') \otimes c = 0$  iff  
 $(a \otimes b \otimes c) \oplus (a \otimes b' \otimes c) = \0$.
\end{compactitem}

We note that zero-distributivity implies the conjunction of zero-right-distributivity and the analogously defined zero-left-distributivity, but the converse does not hold, due to absence of distributivity.

Clearly, we have the implications
\begin{center}
$\B$ positive \ $\Rightarrow$ \ $\B$ zero-distributive \ $\Rightarrow$ \ $\B$ zero-right-distributive.
\end{center}

Next, we will give examples for these classes of strong bimonoids and show that in general the above implications cannot be reversed. A reader interested in the automata-theoretic results may continue directly with Section \ref{sec:wsa} or \ref{sec:wta}. 

There are many well-known examples of positive semirings. We give only a few examples of positive strong bimonoids some of which are not semirings.

\begin{example}\rm  
\begin{enumerate}
\item The Boolean semiring $\Boole = (\{0,1\},\vee,\wedge,0,1)$,
the strong bimonoid $\Nat_{+,\min}  = (\mathbb{N}_\infty,+,\min,0,\infty)$, and
the plus-plus strong bimonoid $\Nat_{+,+} = (\mathbb{N}_\0,\oplus,+,\0,0)$ from Example \ref{ex:strbim} are positive.
\item Every bounded lattice $\sfL = (L, \vee, \wedge, \0, \1)$  is zero-sum-free. If   $\sfL$  contains an element  $a \in L$  such that  
$\0 < a \leq x$ for each $x \in L$ with  $x > \0$, then  $\sfL$  is a positive strong bimonoid. 
\item Near-rings and related structures provide further typical examples, cf. \cite{pil77}.
E.g., the polynomials  $(\mathbb{N}[x],\oplus,\circ,\0,\1)$ over a variable $x$ and with coefficients (e.g.) in $\mathbb{N}$, 
where $\oplus$ is the usual addition of polynomials,
multiplication is given by composition, $\0$ is the zero polynomial and  $\1 = x$,
form a positive strong bimonoid which is right-distributive but not left-distributive: 
for  $p, q, r \in \mathbb{N}[x]$ clearly we have $(p \oplus q) \circ r = p \circ r \oplus q \circ r$, 
but in general $p \circ (q \oplus r) \neq p \circ q \oplus p \circ r$.
\end{enumerate}
\end{example}

As is well-known, there are many natural examples of zero-sum-free semirings (which trivially are zero-distributive) which are not positive.

\begin{example}\rm 
\begin{enumerate}
\item The semiring $(\mathbb{N} \times \mathbb{N},+,\cdot,(0,0),(1,1))$ with componentwise addition and multiplication is a zero-sum-free commutative semiring which is not zero-divisor-free. The semiring $\mathbb{N}^{n\times n}$ of $n\times n$-matrices with the usual addition and multiplication of matrices is zero-sum-free but not zero-divisor-free if  $n \geq 2$.
\item Let  $\B = (B, \vee, \wedge, 0, 1)$  be any Boolean algebra with  $|B| \geq 4$. Then $\B$ is a zero-sum-free commutative semiring which is not positive. 
\end{enumerate}
\end{example}

The following easy characterizations of when a strong bimonoid is zero-right-distributive and zero-sum-free, respectively zero-distributive and zero-sum-free, will be very useful subsequently. The proof is straightforward by checking the definitions.

\begin{observation}\rm \label{obs:zfs-and-zero-distr} Let  $\B$ be a strong bimonoid.
  \begin{compactenum}
  \item[(1)]  $\B$  is zero-right-distributive and zero-sum-free if and only if 
 for all  $a, b, c \in B$  we have:\\
  $(a \oplus b) \otimes c = 0$  iff  $a \otimes c = b \otimes c = \0$. 
  \item[(2)] $\B$  is zero-distributive and zero-sum-free if and only if 
 for all  $a, b, b', c \in B$  we have:\\
  $a \otimes (b\oplus b') \otimes c = 0$  iff  $a \otimes b \otimes c = a \otimes b' \otimes c = \0$. 
  \end{compactenum}
\end{observation}

We also note that the notion of zero-distributivity is well-known and has been much investigated in lattice theory.
A lattice  $\sfL = (L,\vee,\wedge,\0)$  with smallest element  $\0$  is called
\emph{meet-semidistributive at $\0$}, if whenever  $a,b,c \in L$  with  $a \wedge c = b \wedge c = \0$,
then also  $(a \vee b) \wedge c = \0$ (cf., e.g., \cite{adamarmcknatzen06}, p.112). (This condition can be verbally expressed by saying: "If two elements are disjoint to a third element, then their join is also disjoint to this third element.")
The lattice  $\sfL$  is said to be \emph{pseudo-complemented}, 
if for each  $c \in L$  there is a greatest element  $c^* \in L$  such that  $c \wedge c^* = \0$.
Clearly, each pseudo-complemented lattice is meet-semidistributive at $0$, and if $L$  is finite, the converse is also true. Then we have:

\begin{observation} \rm \label{obs:meet-semidistributive-lattice} 
Let  $\sfL = (L,\vee,\wedge,\0, \1)$  be a bounded lattice. 
Then  $\sfL$  is a  zero-distributive strong bimonoid if and only if  $\sfL$  is meet-semidistributive at  $0$.
In particular, each pseudo-complemented lattice is a zero-distributive zero-sum-free strong bimonoid.  
\end{observation}

\begin{proof}  Note that since $a \leq a \vee b$, clearly
$(a \vee b) \wedge c = \0$  implies $a \wedge c = b \wedge c = \0$, for any  $a,b,c \in L$.
Then the defining properties easily translate into each other.
\end{proof}

Next we give two examples of classical bounded lattices, the pentagon and the hexagon (cf. \cite[Fig.~2, p.~14]{gra03}), which are easily seen to be meet-semidistributive at $0$, hence by Observation \ref{obs:meet-semidistributive-lattice} are zero-distributive strong bimonoids, but are neither positive nor right-distributive. These lattices can be easily enlarged by further elements, showing that they are the smallest elements of a class of lattices with the described properties, cf. \cite{riv76, ste91}. 
Hence these lattices satisfy the assumptions of Theorems \ref{thm:wsa-equivalence}(1) and \ref{thm:bi-strongly-zsf-equiv-equ-supp-b}(1), without being positive or right-distributive.

\begin{figure}
\begin{center}
 \centering
  \begin{tikzpicture}[scale=0.75, every node/.style={transform shape,scale=1/0.75},
					mycircle/.style={circle,draw,inner sep=0.05cm}]
\node[mycircle] (a5) {};
\node[mycircle, below= 1.25cm of a5] (b5) {};
\node[mycircle, label={$1$}, above right= 0.75cm and 1cm of a5] (i5) {};
\node[mycircle, label=below:{$0$}, below right= 0.75cm and 1cm of b5] (o5) {};
\node[mycircle, label=right:{}, right= 1.5cm of $(i5)!0.5!(o5)$] (c5) {};
\draw (a5) -- (b5) -- (o5) -- (c5) -- (i5) -- (a5);

\node[mycircle,xshift=5cm] (a6) {};
\node[mycircle, below= 1.25cm of a6] (b6) {};
\node[mycircle, label={$1$}, above right= 0.75cm and 1cm of a6] (i6) {};
\node[mycircle, label=below:{$0$}, below right= 0.75cm and 1cm of b6] (o6) {};
\node[mycircle, below right= 0.75cm and 1cm of i6] (c6) {};
\node[mycircle, above right = 0.75cm and 1cm of o6] (d6) {};
\node[mycircle, right = 1cm of $(a6)!0.5!(b6)$] (e6) {};
\draw (a6) -- (b6) -- (o6) -- (d6) -- (c6) -- (i6) -- (a6);
\draw (i6) -- (e6) -- (b6);
\draw (e6) -- (d6);
 
\end{tikzpicture}
\end{center}
\caption{\label{fig:pentagon-hexagon} Two bounded lattices, the pentagon and the hexagon.}
\end{figure}

\begin{example}\rm \label{pentagon-hexagon}
The two lattices of Figure \ref{fig:pentagon-hexagon} are zero-distributive zero-sum-free commutative strong bimonoids, but are neither positive nor right-distributive.
\end{example}

The following gives a sufficient condition for a zero-right-distributive zero-sum-free strong bimonoid to be positive.

\begin{observation}\rm \label{obs:zfs-implies-positive} \begin{compactenum}
  \item[(1)] Let  $\B$ be a zero-right-distributive zero-sum-free strong bimonoid such that for each $b\in B$ with  $b \neq \0$ there exists  $a \in B$ with  $b = a \oplus \1$. Then  $\B$ is positive.
  \item[(2)] Let  $\otimes$  be a multiplication operation on  $\mathbb{N}$  such that  $(\mathbb{N},+,\otimes,0,1)$  is a zero-right-distributive strong bimonoid. Then $(\mathbb{N},+,\otimes,0,1)$ is positive.
\end{compactenum}
\end{observation}

\begin{proof} Ad (1). Let  $b,c \in B \setminus \{\0\}$. Choose  $a \in B$ with  $b = a \oplus \1$. Then $b \otimes c = (a \oplus \1) \otimes c \neq \0$ 
by Observation \ref{obs:zfs-and-zero-distr}(1)
since $\1 \otimes c \neq \0$.

Ad (2). Immediate by (1).
\end{proof}

For commutative and for left-distributive strong bimonoids, the two conditions of zero-right-distributivity resp. zero-distributivity clearly coincide:

\begin{observation} \rm \label{obs:com-zero-right-distr-implies-zero-distr} A zero-right-distributive strong bimonoid is zero-distributive if it is commutative or left-distributive.
\end{observation}

We note that left-distributive strong bimonoids naturally arise in the theory of near-rings and near-semirings, cf., e.g., \cite{bla53, pil77}.

Now we show that there are zero-sum-free and right-distributive strong bimonoids which are not zero-distributive.

\begin{example}\rm \label{zfs-not-bi-zfs}
There is a zero-sum-free right-distributive strong bimonoid $\B =(B,\oplus,\otimes,\0,\1)$
which is not zero-distributive.
\end{example}

\begin{proof} We use a standard example from the theory of near-rings and near-semirings (cf., e.g., \cite{bla53}; also cf. \cite[Ex.~2.7.15(5)]{fulvog26}) Let $(C,+,0)$ be a commutative monoid. We consider the set  $[C \to C]$ of all mappings
$f: C \to C$  such that  $f(0) = 0$, together with pointwise addition and with composition of mappings; that is,
$(f \oplus g)(c) = f(c) + g(c)$ and $f \circ g(c) = f(g(c))$  for all  $f, g \in [C \to C]$ and $c \in C$.
Also  $\0(c) = 0$ and $ \1(c) = c$ for all $c \in C$. Then  $\B = ([C \to C], \oplus, \circ, \0, \1)$ is a right-distributive strong bimonoid, but in general not left-distributive. 

Now if  $(C,+,0)$  is a zero-sum-free commutative monoid, that is, $a + b = 0$ implies  $a = b = 0$ for all $a, b \in C$,
then the above strong bimonoid  $[C \to C]$ is also zero-sum-free, as is easy to check, but often not zero-distributive. For instance, let  $(C,+,0) = (\mathbb{N},+,0)$ and define  $g: C \to C$  by letting  
$g(2n) = 0$  and  $g(2n+1) = 2n+1$ for each  $n \in \mathbb{N}$. Then $g \circ (g + g) = \0$ but  $g \circ g \neq \0$.
\end{proof}

We note that zero-sum-free right-distributive strong bimonoids which are not semirings can also be obtained, 
similarly as in Example \ref{zfs-not-bi-zfs}, from spaces of non-linear operators on the positive cones of ordered vector spaces occurring, e.g., in Mathematical Physics, cf. \cite[p.3842]{sch97}.

Next we show that there are zero-sum-free and right-distributive strong bimonoids which are not zero-distributive but, nevertheless, satisfy one of the two implications in the definition of zero-distributivity.

\begin{example}\rm  There is a zero-sum-free right-distributive strong bimonoid $\B =(B,\oplus,\otimes,\0,\1)$  satisfying\\[1mm]
\hspace*{10mm}
  $a \otimes (b \oplus b') \otimes c \neq \0  \ \text{ implies } \ a \otimes b \otimes c \neq \0 \text{ or } a \otimes b' \otimes c \neq \0, \ \text{ for all }   a,b,b',c \in B,$ \\[1mm]
  but $\B$  is not zero-distributive. In particular, $\B$ is zero-right-distributive but not zero-distributive.
\end{example}

\begin{proof}  Let  $B = \mathbb{N}\langle x \rangle$ be the set of all polynomials over  $\mathbb{N}$. We let  $\oplus$  be the usual addition of polynomials. We define the multiplication  $p \otimes q$ of two polynomials $p, q \in B$  as follows:
\begin{compactitem}
\item $p \otimes q = p \cdot q$, \ the usual product, if $q$  is a monome; thus, if $p = \sum_{i = 0,...,n} a_i x^i$ and $q = bx^j$  for some  $b \in \mathbb{N}$  and  $j \geq 0$, then  $p \otimes q = \sum_{i = 0,...,n} a_i b x^{i+j}$;
  \item  $p \otimes q = p(0) \cdot q$ \  if $q$ is not a monome, that is, $q = \sum_{k = 0,...,n} b_k x^k$  with  $n \geq 1$  and  $b_i, b_j \neq 0$  for some  $i,j \in [n]$ with  $i \neq j$; in particular, note that if $p = \sum_{i = 1,...,n} a_i x^i$, then  $p \otimes q = \0$.
     \end{compactitem}

Clearly, this multiplication $\otimes$ satisfies the strong bimonoid axioms for  $\0$ and $\1$. We show that $\otimes$ is associative. Let  $p,q,r \in B$. We claim that  $(p \otimes q) \otimes r = p \otimes (q \otimes r)$, and we may assume that  $p,q,r$ are different from $\0$ and $\1$.

First, if $q$ and $r$ are monomes, our claim is clear. Secondly, let  $q$ be a monome, but $r$ not be a monome. Then $(p \otimes q) \otimes r = (p\cdot q)(0) \cdot r = p(0) \cdot q(0) \cdot r$ and $p \otimes (q \otimes r) = p \otimes (q(0) \cdot r)$. Since  $q$ is a monome with $q \neq \0$, we have  $q(0) \neq \0$ and $q(0) \cdot r$ is not a monome. Hence $p \otimes (q(0) \cdot r) = p(0) \cdot (q(0) \cdot r)$ showing our claim.
Thirdly, let $q$ not be a monome, but let $r$ be a monome. Then $(p \otimes q) \otimes r = p(0) \cdot q \cdot r$. Since $r \neq \0$, the product  $q \cdot r$ is not a monome. Hence we obtain $p \otimes (q \otimes r) = p \otimes (q \cdot r) = p(0) \cdot q \cdot r$ as needed. Finally, assume that  $q$ and $r$ are not monomes. Then $(p \otimes q) \otimes r = (p(0) \cdot q)(0) \cdot r = p(0) \cdot q(0) \cdot r$ and $p \otimes (q \otimes r) = p \otimes (q(0) \cdot r)$. If  $q(0) = \0$, both required expressions are $\0$. Therefore we may assume that  $q(0) \neq \0$. Then  $q(0) \cdot r$ is not a monome. Thus $p \otimes (q(0) \cdot r) = p(0) \cdot (q(0) \cdot r)$ as needed. Hence $\otimes$ is associative. 

Clearly, $\otimes$ is right-distributive over $\oplus$, so $\B = (B,\oplus,\otimes,\0,\1)$ is a zero-sum-free right-distributive strong bimonoid. 

First, assume that  $a,b,b' \in B$ with $a \otimes (b \oplus b') \neq \0$. We claim that then  $a \otimes b \neq \0$ or $a \otimes b' \neq \0$. We have  $a \neq \0$  and $b \oplus b' \neq \0$. We may assume that  $b \neq \0$.
Clearly, if $b$ is a monome, then $a \otimes b$ is the usual product and we have $a \otimes b \neq \0$. Next, assume that $b$  is not a monome. Clearly, also  $b \oplus b'$ is not a monome.
Then $a \otimes (b \oplus b') = a(0) \cdot (b \oplus b')$, thus  $a(0) \neq \0$. Hence we have  $a \otimes b = a(0) \cdot b \neq \0$, showing our claim.

Now let $a,b,b',c \in B$ with $a \otimes (b \oplus b') \otimes c \neq \0$. We claim that $a \otimes b \otimes c \neq \0$ or $a \otimes b' \otimes c \neq \0$. We have
$a \otimes (b \oplus b') \otimes c = a \otimes (b \otimes c \oplus b' \otimes c)$. By what we showed above, we directly obtain our claim.

We have $x \otimes (1 \oplus x) = \0$, showing that  $\B$ is not zero-distributive.
\end{proof}

\begin{example}\rm  There is a zero-sum-free right-distributive strong bimonoid $\B =(B,\oplus,\otimes,\0,\1)$  satisfying\\[1mm] 
  \hspace*{10mm} $a \otimes b \otimes c \neq \0    \ \text{ implies } \ a \otimes (b \oplus b') \otimes c \neq \0 \ \text{ for all }   a,b,b',c \in B,$\\[1mm]
  but $\B$  is not zero-distributive.
\end{example}

\begin{proof} (Sketch)
Let  $(B(X), \oplus', \otimes', \0', \1')$  be the free right-distributive strong bimonoid having a non-empty $X$  as its set of free generators. This can be viewed as the set of terms using the elements of  $X$  and  $\0', \1'$ and the operations  $\oplus', \otimes'$ such that the axioms of right-distributive strong bimonoids are satisfied, cf. \cite{drofultepvog24, droful26}.  Let  $\sim$ be the congruence defined by putting  $x \otimes' y \sim \0'$  for all  $x,y \in X$, and let  $\B = B(X)/_\sim$ be the factor algebra of $B(X)$ modulo $\sim$. We write  $\B = (B, \oplus, \otimes, \0, \1)$.
Then  $\B$  is a zero-sum-free right-distributive strong bimonoid. 

Since  $\B$ is right-distributive, to show the required implication it suffices to consider the case $c = \1$.
Now if  $a,b,b' \in B$ with  $a \otimes (b \oplus b') = \0$, we claim that $a \otimes b =  a \otimes b' = \0$. We may assume that  $a \neq \0$ and, since $\B$ is zero-sum-free, that $b \oplus b' \neq \0$. The argument proceeds through an analysis of the term structure of $a$ and
$b \oplus b'$, employing right-distributivity and the definition of $\sim$. As illustration, assume that $a = \bigoplus_{i \in I} [x_i]_\sim$ for some finite collection  $\{x_i | i \in I \} \subseteq X$  and  $b \oplus b' = [y]_\sim$  for some  $y \in X$. Then  $b = [y]_\sim, b' = \0$ or viceversa. In the first case, by right-distributivity,  $a \otimes b = \bigoplus_{i \in I} [x_i \otimes' y]_\sim = \0$  and trivially  $a \otimes b' = \0$ (and similarly in the other case), showing our claim in this case. The general case can be reduced to this subcase. 

For  $x \in X$ we have  $[x]_\sim \otimes [x]_\sim = \0$, 
but  $[x]_\sim \otimes ([x]_\sim \oplus [x]_\sim) = [x \otimes' (x \oplus' x)]_\sim \neq \0$, showing that  $\B$  is not zero-distributive. 
\end{proof}

Finally, we show that there are commutative zero-sum-free strong bimonoids which satisfy either one of the implications of the definition of zero-right-distributivity, but not the other one.

\begin{example}\rm  There are commutative zero-sum-free strong bimonoids $\B =(B,\oplus,\otimes,\0,\1)$  and  $\B' =(B',\oplus',\otimes',\0',\1')$ satisfying, respectively, 
\begin{compactenum}
  \item[(1)]  $a \otimes c \neq \0 \ \text{ implies } \ (a \oplus b) \otimes c \neq \0, \ \text{ for all }   a,b,c \in B,$ 
\item[(2)]  $(a \oplus' b) \otimes' c \neq \0 \ \text{ implies } \ 
a \otimes' c \neq \0 \ \text{or} \ b \otimes' c \neq \0, \ \text{for all } a,b,c \in B',$ 
     \end{compactenum}
but both of them are not zero-right-distributive.
\end{example}
\begin{proof} Let $B = \{ \0, \1, 2, 3 \}$.   We define the operations  $\oplus$  and  $\otimes$ of the commutative strong bimonoid  $\B = (B,\oplus,\otimes, \0, \1)$
such that\\[1mm]
\hspace*{10mm}$\1 \oplus \1 = \1, \ \ 2 \oplus j = 3 \oplus j = 3$ for each $j \in \{ \1, 2, 3 \}$ and  $2 \otimes 2 = \0, \ 2 \otimes 3 = 2, \ 3 \otimes 3 = 3$.

Then $\B$  is commutative and zero-sum-free and satisfies (1) but not the corresponding condition of (2), since  $(2 \oplus 2) \otimes 2 = 3 \otimes 2 = 2 \neq \0$, but  $2 \otimes 2 = \0$. 

Secondly, let  $B' = \{ \0', \1', 2' \}$. We define the operations  $\oplus'$  and  $\otimes'$ of the commutative strong bimonoid  $\B' = (B',\oplus',\otimes', \0', \1')$
such that\\[1mm]
\hspace*{10mm} $\1' \oplus' \1' = \1' \oplus' 2' = 2' \oplus' 2 = 2'$ and $2' \otimes 2' = \0'$.

Then $\B'$  is commutative and zero-sum-free and satisfies (2) but not the corresponding condition of (1), since  $(\1' \oplus' \1') \otimes' 2' = \0'$. 
\end{proof}

\section{Weighted automata over words}
\label{sec:wsa}
 
We recall the concept of weighted automata over strong bimonoids from \cite{drostuvog10,cirdroignvog10}.

Let $\Gamma$ be an alphabet and $\B = (B,\oplus,\otimes,\0,\1)$ a strong bimonoid.  A \emph{weighted automaton over $\Gamma$ and $\B$} is a tuple $\cA=(Q,I,\mu,F)$ where $Q$ is a finite non-empty set, $I \in B^Q$ (\emph{initial weight vector}), $F \in B^Q$ (\emph{final weight vector}), and 
$\mu:Q \times \Gamma \times Q \to B$ (\emph{transition weight mapping}).
Here for  $a \in \Gamma$, in slight abuse of notation, the  $Q \times Q$-matrix over  $B$  denoted by  $\mu(a)$  is defined by letting  $\mu(a)_{p,q} = \mu(p,a,q)$  $(p,q \in Q)$.

\paragraph{Run semantics.} The \emph{run semantics of $\cA$} is the weighted language $\runsem{\cA}: \Gamma^* \to B$ defined for each $w = a_1 \cdots a_n$ in $\Gamma^*$ (with $n \in \mathbb{N}$ and $a_i \in \Gamma$) by letting
\[
\runsem{\cA}(w) = \bigoplus_{P \in Q^{n+1}} \wt_\cA(w,P)
  \]
  where for every $P = (q_0,\ldots, q_{n})$ in $Q^{n+1}$ (-- $P$ is called a \emph{run} --), the \emph{weight of $P$  in  $\cA$ for $w$} is defined as
  \[
\wt_\cA(w,P) = I_{q_0} \otimes \mu(q_0,a_0,q_1) \otimes \cdots \otimes \mu(q_{n-1},a_n,q_n) \otimes F_{q_n} \enspace.
    \]
    Thus, in particular, $\wt_\cA(\varepsilon,q) = I_q \otimes F_q$. The set $\supp(\runsem{\cA})$ is called the \emph{run support of $\cA$}.

\paragraph{Initial algebra semantics.}    The \emph{initial algebra semantics of $\cA$} is the weighted language $\initialsem{\cA}: \Gamma^* \to B$ defined for each $w$ in $\Gamma^*$ by letting
    \[
\initialsem{\cA}(w) = \bigoplus_{q \in Q} \h_\cA(w)_q \otimes F_q
\]
where $\h_\cA: \Gamma^* \to B^Q$ is defined for each $w = a_1 \cdots a_n$ in $\Gamma^*$ (with $n \in \mathbb{N}$ and $a_i \in \Gamma$) by letting
\[
\h_\cA(w) = (\cdots ((I \otimes \mu(a_1)) \otimes \mu(a_2))\cdots  \otimes \mu(a_{n-1})) \otimes \mu(a_n)
\]
using vector multiplication. 
Thus, in particular, $\h_\cA(\varepsilon) = I$.
The set $\supp(\initialsem{\cA})$ is called the \emph{initial algebra support of $\cA$}.

If $\cA$ is clear from the context, then we drop $\cA$ from $\wt_\cA$ and $\h_\cA$ and simply write $\wt$ and $\h$, respectively.

For the convenience of the reader, we point out why this function  $\initialsem{\cA}$ actually constitutes an initial algebra semantics as described in \cite{gogthawagwri77}. This exploits  the concepts of $\Sigma$-algebras and the initiality of $\Gamma^*$ . 

More precisely, each symbol $a \in \Gamma$ is interpreted as unary operation on $\Gamma^*$, which appends $a$ to its argument. Moreover, there is an extra symbol, say, $e \not\in \Gamma$, which is interpreted as nullary operation on $\Gamma^*$ yielding the empty word $\varepsilon$. (Thus we are not dealing with the monoid structure of $\Gamma^*$.) Then $\Gamma^*$ is initial in the class of all algebras with signature $\Gamma \cup \{e\}$, i.e., for each algebra $\cB$ with signature $\Gamma \cup \{e\}$ there exists exactly one homomorphism from $\Gamma^*$ to $\cB$.

A weighted automaton $\cA$ over $\Gamma$ and $\B$ with state set $Q$ determines another particular algebra with signature $\Gamma \cup \{e\}$. It has the carrier set $B^Q$ and, as seen above, it interprets $e$  as the initial weight vector of $\cA$ and each input symbol $a \in \Gamma$ as a unary operation on $B^Q$, operating by multiplication with the matrix  $\mu(a)$ on $B^Q$.  Due to initiality, there is a unique homomorphism $\h$ from $\Gamma^*$ to $B^Q$. This homomorphism $\h$ is precisely the mapping $\h_\cA$ defined above by induction. Finally, the image of an input structure under $\h_\cA$ is multiplied with the final weight vector of $\cA$. This results in the initial algebra semantics of $\cA$. For details, we refer the reader to \cite[p.115]{fulvog26}.

 \paragraph{Comparison of semantics.}
 For each weighted automaton over $\Gamma$ and $\B$, its initial algebra semantics coincides with its run semantics if $\B$ is a semiring. Even the following characterization holds.

\begin{theorem}\rm \cite[Lm.~4]{drostuvog10} \label{thm:wsa-semiring-run=initial} Let $\Gamma$ be an alphabet and  $\B$ be a strong bimonoid. The following two statements are equivalent:
\begin{compactenum}
\item[(1)] $\B$ is right-distributive.
\item[(2)] For every  weighted automaton $\cA$ over  $\Gamma$  and  $\B$, we have $\runsem{\cA} = \initialsem{\cA}$.
\end{compactenum}
\end{theorem}

The following result summarizes the main known sufficient conditions for a strong bimonoid  $\B$ ensuring that for each weighted automaton  $\cA$ over an alphabet  $\Gamma$ and $\B$, the run and the initial algebra supports of $\cA$ coincide; for the case of positive strong bimonoids cf. \cite[Thm.~4.1]{ghovog23};
the case that  $\B$ is right-distributive is immediate by Theorem \ref{thm:wsa-semiring-run=initial} because the two semantics' of $\cA$ coincide.

\begin{theorem} \label{thm:sufficient-cond-wsa-equality-supp}
  Let $\Gamma$ be an alphabet and  $\B$ be a strong bimonoid. If  $\B$ is positive or right-distributive, then for each weighted automaton $\cA$ over  $\Gamma$  and  $\B$, we have $\supp(\runsem{\cA}) = \supp(\initialsem{\cA})$. 
\end{theorem}


\subsection{Support results for weighted automata}
\label{sec:support-result-for-wsa}

The main goal of this section will be the proof of the following result. It characterizes Property~\eqref{eq:equality-of-supports}, given in the Introduction, for weighted automata over words and zero-sum-free strong bimonoids in terms of zero-right-distributivity. 

\begin{theorem-rect} \label{thm:wsa-equivalence} Let $\Gamma$ be an alphabet and $\B$ be a zero-sum-free strong bimonoid. Then the following statements are equivalent.
  \begin{compactenum}
  \item[(1)] $\B$ is zero-right-distributive.
  \item[(2)] For each weighted automaton $\cA$ over  $\Gamma$  and  $\B$, we have $\supp(\runsem{\cA}) = \supp(\initialsem{\cA})$.
    \end{compactenum}
  \end{theorem-rect}

Note the formal analogy of Theorem~\ref{thm:wsa-equivalence} to Theorem~\ref{thm:wsa-semiring-run=initial}: in condition (2), we have weakened the equality of the two behaviors (semantics) to the equality of the two behaviors with respect to $0$ (the supports), and this corresponds precisely to a weakening of right-distributivity of the given strong bimonoid to zero-right-distributivity.
In fact, for each of the two inclusions of condition (2) of Theorem \ref{thm:wsa-equivalence}, we will prove a characterization (cf. Theorems~\ref{thm:run-=-init} and \ref{thm:init-=-run}).
Recall that the run semantics is computed, roughly, by taking sums of products of values, whereas the initial algebra semantics is computed in a more involved way by iterating sums and products. Condition (2) of the subsequent Theorem \ref{thm:run-=-init} says that whenever the run semantics produces a non-zero value then so does the initial algebra semantics. Theorem \ref{thm:run-=-init} characterizes this by a simple condition on the strong bimonoid $\B$ which means that non-zero products are preserved by sum extensions. In Theorem \ref{thm:init-=-run} the converse condition on the run and initial algebra semantics is characterized by the ``converse'' condition on the strong bimonoid $\B$ saying that sums in non-zero products can be cancelled. These conditions can also be seen as one implication of the equivalence in the definition of zero-right-distributivity.
Theorem \ref{thm:wsa-equivalence} will then be an immediate consequence of Theorems \ref{thm:run-=-init} and \ref{thm:init-=-run}.

In the proof of the subsequent two results, we will employ the following simple weighted automaton. 

\begin{definition}\rm \label{ex:special-wsa}
We fix  $a,b,c \in B$  and  $\gamma \in \Gamma$.
We define the weighted automaton $\cA = (Q, I, \mu, F)$ over $\Gamma$ and $\B$ by letting
carrier set\begin{compactitem}
\item $Q=\{p,q,r\}$, 
\item $I_p=a$, $I_q=b$, $I_r = \0$,
\item $\mu(p,\gamma,r)=\mu(q,\gamma,r)=\1$ and
  $\mu(q',\gamma,r') = \mu(r,\gamma,r) = \0$ 
for all  $q' \in Q$ and $r' \in \{p, q\}$, and
    $\mu(q',\gamma',q'') = \0$  for all  $q',q'' \in Q$ and $\gamma' \in \Gamma$ with $\gamma' \neq \gamma$.

\item $F_p=F_q=\0$ and  $F_r=c$.
 \end{compactitem}

 Then 
\begin{eqnarray}
\runsem{\cA}(\gamma) = a \otimes c \oplus b \otimes c \ \text{ and } \
\initialsem{\cA}(\gamma) =  (a \oplus b) \otimes c \label{equ:wsa-specialtree-sem}\enspace.
\end{eqnarray}

Since $\supp(\runsem{\cA}) \subseteq  \{\gamma\}$, $\supp(\initialsem{\cA}) \subseteq  \{\gamma\}$, and $\runsem{\cA}(\varepsilon) = \initialsem{\cA}(\varepsilon)= \0$, we also have 
\begin{eqnarray}
\im(\runsem{\cA}) = \{\0, a \otimes c \oplus b \otimes c\} \ \text{ and } \
\im(\initialsem{\cA}) = \{\0, (a \oplus b) \otimes c\}\enspace. \label{equ:wsa-im-sem}
\end{eqnarray}
\hfill $\Box$
\end{definition}

\begin{theorem}\rm \label{thm:run-=-init}  Let  $\Gamma$  be an alphabet and $\B$ be a zero-sum-free strong bimonoid. Then the following two statements are equivalent.
  \begin{compactenum}
    \item[(1)]  
    $a \otimes c \neq \0 \ \text{ implies } \ (a \oplus b) \otimes c \neq \0, \ \text{ for all   $a,b,c \in B$.}$
    
  \item[(2)] For each weighted automaton $\cA$ over  $\Gamma$  and  $\B$, we have $\supp(\runsem{\cA}) \subseteq \supp(\initialsem{\cA})$.
    \end{compactenum}
  \end{theorem}

\begin{proof} $(1)\Rightarrow$(2): 
Let $\cA = (Q,I,\mu,F)$ be a weighted automaton over  $\Gamma$  and  $\B$ and let $w \in \supp(\runsem{\cA})$.
Let $w = a_1 \cdots a_n$ for some $n \in \mathbb{N}$ and $a_i \in \Gamma$ ($i \in [n]$).

Since $\runsem{\cA}(w) = \bigoplus_{P \in Q^{n+1}} \wt(w,P) \not= \0$, there exists a run
$P \in Q^{n+1}$  such that $\wt(w,P) \not= \0$. Let $P = q_0 \cdots q_n$ with $q_i \in Q$. Then 
\begin{equation}\label{equ:run-not-0}
 \wt(w,P) =  I_{q_0} \otimes \mu(q_0, a_1, q_1) \otimes ... \otimes \mu(q_{n-1}, a_n, q_n) \otimes F_{q_n} \neq \0\enspace.
\end{equation}

By bounded induction in $\mathbb{N}$, we prove that the following statement holds.
\begin{equation}\label{equ:h-not-equal-0}
\text{For each $i \in [0,n]$, we have} \ \h(a_1 \cdots a_i)_{q_i} \otimes \Big(\bigotimes_{j=i}^{n-1} \mu(q_j,a_{j+1},q_{j+1})\Big) \otimes F_{q_n}\not= \0
\end{equation}

First, by definition, we have $\h(\varepsilon)_{q_0} = I_{q_0}$. Hence equation \eqref{equ:run-not-0} implies equation \eqref{equ:h-not-equal-0} for $i=0$.

Now let $i \in [0,n-1]$ and assume that equation \eqref{equ:h-not-equal-0} holds. Then we have:
\begingroup
\allowdisplaybreaks
\begin{align*}
  &   \h(a_1 \cdots a_i)_{q_i} \otimes \Big(\bigotimes_{j=i}^{n-1} \mu(q_j,a_{j+1},q_{j+1})\Big) \otimes F_{q_n}\not= \0\\
\Leftrightarrow \ \ &   \Big(\h(a_1 \cdots a_i)_{q_i} \otimes \mu(q_i,a_{i+1},q_{i+1})\Big) \otimes \Big(\bigotimes_{j=i+1}^{n-1} \mu(q_j,a_{j+1},q_{j+1})\Big) \otimes F_{q_n}\not= \0 \tag{because $i+1 \le n$}\\
  \Rightarrow \ \ &   \Big(\h(a_1 \cdots a_i)_{q_i} \otimes \mu(q_i,a_{i+1},q_{i+1}) \oplus
                    \bigoplus_{q \in Q\setminus\{q_i\}}\h(a_1 \cdots a_i)_{q} \otimes \mu(q,a_{i+1},q_{i+1}) \Big)\\
  & \hspace*{70mm} \otimes \Big(\bigotimes_{j=i+1}^{n-1} \mu(q_j,a_{j+1},q_{j+1})\Big) \otimes F_{q_n}\not= \0 \tag{by Statement (1) of the theorem}\\
    \Leftrightarrow \ \ &   \Big(\bigoplus_{q \in Q}\h(a_1 \cdots a_i)_{q} \otimes \mu(q,a_{i+1},q_{i+1}) \Big)
                          \otimes \Big(\bigotimes_{j=i+1}^{n-1} \mu(q_j,a_{j+1},q_{j+1})\Big) \otimes F_{q_n}\not= \0\\
  \Leftrightarrow \ \ &  \h(a_1 \cdots a_{i+1})_{q_{i+1}} \otimes \Big(\bigotimes_{j=i+1}^{n-1} \mu(q_j,a_{j+1},q_{j+1})\Big) \otimes F_{q_n}\not= \0 \tag{by definition of $\h(a_1 \cdots a_{i+1})_{q_{i+1}}$}
\end{align*}
\endgroup
  This proves equation \eqref{equ:h-not-equal-0} for $i+1$. In particular, for $i=n$, we obtain:
  \(\h(a_1 \cdots a_n)_{q_n} \otimes  F_{q_n}\not= \0\). 
Hence, since  $\B$ is zero-sum-free, we have
\[
\sem{\cA}^{\mathrm{init}}(a_1 \cdots a_n) = \bigoplus_{q \in Q} \h(a_1 \cdots a_n)_q \otimes F_{q} \not=\0
  \]
Thus $w \in \supp(\initialsem{\cA})$.

$(2)\Rightarrow$(1): Let  $a,b,c \in B$ with  $a \otimes c \neq 0$. Consider the weighted automaton  $\cA$  given in Definition~\ref{ex:special-wsa}. Since  $\B$  is zero-sum-free, we have $a \otimes c \oplus b \otimes c \neq \0$ and thus  $\gamma \in \supp(\runsem{\cA})$ by~\eqref{equ:wsa-specialtree-sem}. Then $\gamma \in \supp(\initialsem{\cA})$, showing $(a \oplus b) \otimes c \neq \0$ again  by~\eqref{equ:wsa-specialtree-sem}. 
\end{proof}

The following result characterizes the converse inclusion for the supports.

  \begin{theorem}\rm \label{thm:init-=-run}  Let  $\Gamma$  be an alphabet and $\B$ be a zero-sum-free strong bimonoid. Then the following two statements are equivalent.
  \begin{compactenum}
    \item[(1)]  $(a \oplus b) \otimes c \neq \0$ implies $a \otimes c \neq \0$ or $b \otimes c \neq \0$, for all   $a,b,c \in B$ . 
  \item[(2)] For each weighted automaton $\cA$ over  $\Gamma$  and  $\B$, we have $\supp(\initialsem{\cA}) \subseteq \supp(\runsem{\cA})$.
    \end{compactenum}
\end{theorem}

\begin{proof} $(1)\Rightarrow$(2): 
Let $\cA = (Q,I,\mu,F)$ be a weighted automaton over  $\Gamma$  and  $\B$ and let $w \in \supp(\initialsem{\cA})$.
Let $w = a_1 \cdots a_n$ for some $n \in \mathbb{N}$ and $a_i \in \Gamma$ ($i \in [n]$).

Since $\initialsem{\cA}(w) = \bigoplus_{q \in Q} \h(w)_q \otimes F_q \not= \0$, there exists $q \in Q$ such that $\h(w)_q \otimes F_q \not= \0$. Denote this $q$ by $q_n$.

By bounded induction on $\mathbb{N}$, we prove that the following statement holds.
\begin{eqnarray}\label{equ:run-i-not-0}
  &\text{For every $i \in [0,n]$ there exist $q_{n-i},\ldots,q_{n-1} \in Q$ such that }\\
 & \h(a_1 \cdots a_{n-i})_{q_{n-i}} \otimes \big(\bigotimes_{j=n-i}^{n-1} \mu(q_j,a_{j+1},q_{j+1})\big) \otimes F_{q_n} \not= \0 \enspace.  \notag
\end{eqnarray}
We note that, for $i=0$, the list of existential quantifications is empty.

First, let $i=0$. Then $\h(a_1 \cdots a_{n-i})_{q_{n-i}} = \h(w)_{q_n}$ and  $\h(w)_{q_n} \otimes F_{q_n} \not= \0$ by the above.

Next, let $i \in [0,n-1]$ and assume that \eqref{equ:run-i-not-0} holds for $i$, i.e., there exist $q_{n-i},\ldots,q_{n-1} \in Q$ such that
$\h(a_1 \cdots a_{n-i})_{q_{n-i}} \otimes \big(\bigotimes_{j=n-i}^{n-1} \mu(q_j,a_{j+1},q_{j+1})\big) \otimes F_{q_n} \not= \0$. Then we have:
\begin{align*}
&\h(a_1 \cdots a_{n-i})_{q_{n-i}} \otimes \big(\bigotimes_{j=n-i}^{n-1} \mu(q_j,a_{j+1},q_{j+1})\big) \otimes F_{q_n} \not= \0\\
\Leftrightarrow \ \  & \Big( \bigoplus_{q \in Q} \h(a_1 \cdots a_{n-i-1})_q \otimes \mu(q,a_{n-i},q_{n-i}) \Big) \otimes \big(\bigotimes_{j=n-i}^{n-1} \mu(q_j,a_{j+1},q_{j+1})\big) \otimes F_{q_n} \not= \0 \tag{because $i \le n-1$, and hence $n-i \ge 1$}\\
\Rightarrow \ \  & \Big(\h(a_1 \cdots a_{n-i-1})_{q_{n-i-1}} \otimes \mu(q_{n-i-1},a_{n-i},q_{n-i}) \Big) \otimes \big(\bigotimes_{j=n-i}^{n-1} \mu(q_j,a_{j+1},q_{j+1})\big) \otimes F_{q_n} \not= \0 \tag{by Statement (1) of the theorem there exists such a $q_{n-i-1}$}\\
\Leftrightarrow \ \  & \h(a_1 \cdots a_{n-i-1})_{q_{n-i-1}} \otimes \Big(\mu(q_{n-i-1},a_{n-i},q_{n-i}) \otimes \bigotimes_{j=n-i}^{n-1} \mu(q_j,a_{j+1},q_{j+1})\Big) \otimes F_{q_n} \not= \0 \\
\Leftrightarrow \ \  & \h(a_1 \cdots a_{n-(i+1)})_{q_{n-(i+1)}} \otimes \Big(\bigotimes_{j=n-(i+1)}^{n-1} \mu(q_j,a_{j+1},q_{j+1})\Big) \otimes F_{q_n} \not= \0
\end{align*}
This proves equation \eqref{equ:run-i-not-0} for $i+1$. Thus, for $i=n$, there exist 
$q_0,\ldots,q_{n-1} \in Q$ such that
\[
  I_{q_0} \otimes \Big(\bigotimes_{j=0}^{n-1} \mu(q_j,a_{j+1},q_{j+1})\Big) \otimes F_{q_n} \not= \0
\]
taking into account that $I_{q_0} = \h(\varepsilon)_{q_0}$. Thus there exists $r = q_0\cdots q_n$ in $Q^{n+1}$ such that $\wt(w,P) \not= \0$. Since $\B$ is zero-sum-free, we obtain
\[
\runsem{\cA}(w) = \bigoplus_{P \in Q^{n+1}} \wt(w,P) \not= \0 \enspace.
\]
Hence $w \in \supp(\runsem{\cA})$.

$(2)\Rightarrow$(1): Let  $a,b,c \in B$ with $(a \oplus b) \otimes c \neq \0$. Consider the weighted automaton  $\cA$  given in Definition~\ref{ex:special-wsa}. Then $\gamma \in \supp(\initialsem{\cA})$ by~\eqref{equ:wsa-specialtree-sem}. By our assumption we obtain $\gamma \in \supp(\runsem{\cA})$, hence $a \otimes c \oplus b \otimes c \neq \0$ (again by using~\eqref{equ:wsa-specialtree-sem}) and then $a \otimes c \neq \0$ or $b \otimes c \neq \0$. 
\end{proof}

Now the \emph{Proof of Theorem \ref{thm:wsa-equivalence}} is immediate by Theorems  \ref{thm:run-=-init} and \ref{thm:init-=-run}.

    \section{Weighted automata over trees} \label{sec:wta}

    We recall the definitions of weighted automata over trees, for short: weighted tree automata, and their run semantics and initial algebra semantics from \cite[Sec.~3.1]{fulvog26}.
    
Let $\Sigma$ be a ranked alphabet and $\B=(B,\oplus,\otimes,\0,\1)$ a strong bimonoid. A \emph{weighted tree automaton over $\Sigma$ and $\B$} is a tuple $\cA = (Q,\delta,F)$ where 
$Q$ is a finite non-empty set (\emph{states}) such that $Q \cap \Sigma = \emptyset$,
 $\delta=(\delta_k\mid k\in\mathbb{N})$ is a family of mappings $\delta_k: Q^k\times \Sigma^{(k)}\times Q \to B$ (\emph{transition mappings})  where we consider $Q^k$ as set of words over $Q$ of length $k$, and 
 $F: Q \rightarrow B$ is a mapping (\emph{root weight vector}).

\paragraph{Run semantics.}
Let $\xi \in \T_\Sigma$. A \emph{run of $\cA$ on $\xi$} is a mapping $\rho: \pos(\xi) \rightarrow Q$.  The \emph{set of all runs of $\cA$ on $\xi$} is denoted by $\R_\cA(\xi)$.
For every $\rho\in \R_\cA(\xi)$ and $w \in\pos(\xi)$, the \emph{run induced by~$\rho$ at position~$w$}, denoted by $\rho|_w$, is the run in $\R_\cA(\xi|_w)$  defined for every $w'\in\pos(\xi|_w)$ by $\rho|_w(w') = \rho(ww')$. (Cf. Figure \ref{fig:well-founded-order-run} for an example of a tree $\xi \in \T_\Sigma$ and a run $\rho \in \R_\cA(\xi)$, and the run $\rho|_1 \in \R_\cA(\xi|_1)$ induced by $\rho$ at position $w=1$.)

  \begin{figure}
  \centering
\begin{tikzpicture}[scale=0.8, every node/.style={transform shape},
					node distance=0.05cm and 0.05cm,
					level distance= 1.25cm,
					level 1/.style={sibling distance=25mm},
			        level 2/.style={sibling distance=17mm},
					mycircle/.style={draw, circle, inner sep=0mm, minimum height=5.5mm},]

\node (n0) {$\gamma$}
    child {node (n1) {$\sigma$}
      child {node (n2) {$\alpha$}}
      child {node (n3) {$\beta$}} };
  \node[mycircle, right= of n0] (cn0) {$q$};
  \node[mycircle, right= of n1] (cm1) {$p$};
  \node[mycircle, right= of n2] (cn2) {$q$};
  \node[mycircle, right= of n3] (cn3) {$q$};
  \node[above left= 0cm and 0.2cm of n0] (n) {$\xi|_1 \in \T_\Sigma$};
  \node[right= 1.6cm of n] {$\rho|_1 \in \R_{\cA}(\xi|_1)$};

  \node[left= 7cm of n0, yshift=12mm] (m0) {$\sigma$}
 	child {node (mn0) {$\gamma$} 
      child {node (mn1) {$\sigma$}
        child {node (mn2) {$\alpha$}}
        child {node (mn3) {$\beta$}} }}
    child {node (m1) {$\alpha$}};
  \node[mycircle, right= of m0] (cm0) {$q$};
  \node[mycircle, right= of mn0] (cmn0) {$q$};
  \node[mycircle, right= of mn1] (cmn1) {$p$};
  \node[mycircle, right= of mn2] (cmn2) {$q$};
  \node[mycircle, right= of mn3] (cmn3) {$q$};
  \node[mycircle, right= of m1] (cm1) {$p$};
  \node[above left=-0cm and 0.2cm of m0] (m) {$\xi \in \T_\Sigma$};
  \node[right= 1.6cm of m] {$\rho \in \R_{\cA}(\xi)$};

\end{tikzpicture}
\caption{\label{fig:well-founded-order-run} A tree $\xi \in \T_\Sigma$ and a run $\rho \in \R_\cA(\xi)$ (left), and the run $\rho|_1 \in \R_\cA(\xi|_1)$ induced by $\rho$ at position $w=1$ (right), \cite[Fig.~3.2]{fulvog26}.}
  \end{figure}

We define the weight of a run  $\rho$ on $\xi$ as follows.
We put  $\mathrm{TR} = \{(\xi,\rho) \mid \xi \in \T_\Sigma, \rho \in \R_\cA(\xi)\}$ and 
we define the mapping
  \(\wt_\cA: \mathrm{TR} \to B\)
   for every $\xi \in \T_\Sigma$ and $\rho \in \R_\cA(\xi)$ inductively by 
\begin{equation}\label{equ:weight-of-run}
\wt_\cA(\xi,\rho) = \Big( \bigotimes_{i\in [k]} \wt_\cA(\xi|_i,\rho|_i)\Big) \otimes \delta_k\big(\rho(1) \cdots \rho(k),\sigma,\rho(\varepsilon)\big) \enspace,
\end{equation}
where $k$ and $\sigma$ abbreviate $\rk(\xi(\varepsilon))$ and $\xi(\varepsilon)$, respectively. Note that empty products are defined to be $\1$. 
We call $\wt_\cA(\xi,\rho)$ the \emph{weight of $\rho$ by $\cA$ on $\xi$}.

The {\it run semantics of $\cA$}, denoted by $\runsem{\cA}$, is the weighted tree language $\runsem{\cA}:~\T_\Sigma~\rightarrow~B$ such that,  for each $\xi \in \T_\Sigma$, we let
\begin{equation*}
  \runsem{\cA}(\xi) = \bigoplus_{\rho \in \R_\cA(\xi)}\wt_\cA(\xi,\rho) \otimes F_{\rho(\varepsilon)}\enspace. 
  \end{equation*}

  \begin{observation}\rm \label{obs:weight-run-explicit}  For every $\xi \in \T_\Sigma$ and $\rho\in \R_\cA(\xi)$, we have
  \[
\wt_\cA(\xi,\rho) = \bigotimes_{\substack{u \in \pos(\xi)\\\text{traversal in $<_\mathrm{po}$}}} \delta_{\rk(\xi(u))}(\rho(u1) \cdots \rho(u \, \rk(\xi(u))),\xi(u),\rho(u))\enspace.
\]
\end{observation}

If $\cA$ is clear from the context, then we drop $\cA$ from  $\wt_\cA$ and simply write  $\wt$.
The set $\supp(\runsem{\cA})$ is called the \emph{run support of $\cA$}.

\paragraph{Initial algebra semantics.}

We define the mapping $\h_\cA: \T_\Sigma \to B^Q$ inductively as follows.
For every $k \in \mathbb{N}$, $\sigma \in \Sigma^{(k)}$, $\xi_1,\ldots,\xi_k \in \T_\Sigma$, and $q \in Q$  we let
\begin{equation}\label{eq:delta-A-definition}
\h_\cA(\sigma(\xi_1,\dots,\xi_k))_q 
  = \bigoplus_{q_1\cdots q_k \in Q^k} \Big(\bigotimes_{i\in[k]} \h_\cA(\xi_i)_{q_i}\Big) \otimes \delta_k(q_1\cdots q_k,\sigma,q)
\end{equation}
We note that, for each $\alpha \in \Sigma^{(0)}$, we have $\h_\cA(\alpha)_q = \delta_0(\varepsilon,\alpha,q)$, because $Q^0=\{\varepsilon\}$ an$\wt_\cA$ and $\bigotimes_{i\in \emptyset} \h_\cA(\xi_i)_{q_i}=\1$.

The \emph{initial algebra semantics of $\cA$}, denoted by $\initialsem{\cA}$, is the weighted tree language $\initialsem{\cA}: \T_\Sigma \rightarrow B$  defined for every $\xi\in \T_\Sigma$ by

\begin{equation}
  \initialsem{\cA}(\xi) = \bigoplus_{q \in Q} \h_\cA(\xi)_q \otimes F_q\enspace. 
  \end{equation}

If $\cA$ is clear from the context, then we drop $\cA$ from  $\h_\cA$ and simply write  $\h$.
The set $\supp(\initialsem{\cA})$ is called the \emph{initial algebra support of $\cA$}.

Again for the convenience of the reader, we point out why this function  $\initialsem{\cA}$ constitutes an initial algebra semantics as described in \cite{gogthawagwri77}. Here,  the initiality of the $\Sigma$-term algebra  $\T_\Sigma$ is exploited.

In the $\Sigma$-term algebra,  each $k$-ary symbol $\sigma \in \Sigma$ (with $k \ge 0$)  is interpreted as $k$-ary operation on $\T_\Sigma$, which glues $\sigma$ on top of the argument trees. Then $\T_\Sigma$ is  initial in the class of all algebras with signature $\Sigma$, i.e., for each algebra $\cB$ with signature $\Sigma$ there exists exactly one homomorphism from the $\Sigma$-term algebra to $\cB$.

A weighted tree automaton $\cA$ over $\Sigma$ and $\B$ with state set $Q$ determines another  algebra with signature~$\Sigma$. It has the carrier set $B^Q$ and, as seen above, it interprets each input symbol $k$-ary $\sigma \in \Sigma$ as a $k$-ary operation on $B^Q$. Due to initiality, there is a unique homomorphism $\h$ from $\T_\Sigma$ to $B^Q$. Again, this homomorphism $\h$ is precisely the mapping $\h_\cA$ defined by induction above. Finally, the image of an input structure under $\h_\cA$ is multiplied with the final weight vector of $\cA$. This results in the initial algebra semantics of $\cA$. For details, we refer the reader to \cite[p.85]{fulvog26}.

 \paragraph{Comparison of semantics and special cases.}
If $\Sigma$ is trivial, then $\T_\Sigma = \Sigma^{(0)}$ and for each $(\Sigma,\B)$-wta $\cA$ we have $\initialsem{\cA} = \runsem{\cA}$. Therefore subsequently we consider non-trivial ranked alphabets  $\Sigma$. 
 
 For each weighted tree automaton over $\Sigma$ and $\B$, its initial algebra semantics coincides with its run semantics if $\B$ is a semiring \cite[Lm. 4.1.13]{bor04b}. Even the following characterization holds.

\begin{theorem}\rm \label{thm:wta-semiring-runsem=initialsem} \cite[Thm.~4.1]{rad10} \label{thm:wta-semiring-run=initial} Let $\Sigma$ be a non-trivial ranked alphabet, and let $\B$ be a strong bimonoid. The following two statements are equivalent:
\begin{compactenum}
\item[(1)] If $\Sigma$ is monadic, then $\B$ is right-distributive, and if $\Sigma$ is branching, then $\B$ is distributive.
\item[(2)] For each weighted tree automaton $\cA$ over $\Sigma$ and $\B$, we have $\runsem{\cA} = \initialsem{\cA}$.
\end{compactenum}
\end{theorem}

In contrast to string ranked alphabets, a monadic ranked alphabet  may contain several nullary symbols. But we have the following lifting properties.

\begin{observation}\rm \label{obs:restriction-of-wta-to-one-nullary-symbol} Let $\Sigma$ be a ranked alphabet which is non-trivial and monadic. For each $\alpha \in \Sigma^{(0)}$ let $\Sigma_\alpha$ be the string ranked alphabet  $\{\alpha^{(0)}\} \cup \Sigma^{(1)}$. Then
   \begin{compactenum}
    \item[(1)] $\T_\Sigma = \bigcup_{\alpha \in \Sigma^{(0)}} \T_{\Sigma_\alpha}$.
    \end{compactenum}
    Moreover, let $\cA$ be a weighted tree automaton over $\Sigma$ and $\B$ and, for each $\alpha \in \Sigma^{(0)}$, let  $\cA_\alpha$ be the restriction of $\cA$ to the alphabet  $\Sigma_\alpha$. Then the following three statements hold: 
  \begin{compactenum}
    \item[(2)] for each $\xi \in \T_{\Sigma_\alpha}$, we have  $\runsem{\cA}(\xi) = \runsem{\cA_\alpha}(\xi)$ and $\initialsem{\cA}(\xi) = \initialsem{\cA_\alpha}(\xi)$,
    \item[(3)]  $\supp(\runsem{\cA}) = \bigcup_{\alpha \in \Sigma^{(0)}} \supp(\runsem{\cA_\alpha})$ and
      $\supp(\initialsem{\cA}) = \bigcup_{\alpha \in \Sigma^{(0)}} \supp(\initialsem{\cA_\alpha})$, and
    \item[(4)]  $\im(\runsem{\cA}) = \bigcup_{\alpha \in \Sigma^{(0)}} \im(\runsem{\cA_\alpha})$ and
      $\im(\initialsem{\cA}) = \bigcup_{\alpha \in \Sigma^{(0)}} \im(\initialsem{\cA_\alpha})$.
\end{compactenum}
  \end{observation}

 It is well-known that weighted tree automata over string ranked alphabets are equivalent to weighted automata over words (for initial algebra semantics in the unweighted case cf. \cite[p.~55]{gecste84}). This can be seen as follows. Essentially, words are trees over a string ranked alphabet, and vice versa. Formally, given an alphabet $\Gamma$ and  $e \not\in \Gamma$, we define the string ranked alphabet $\Gamma_e$ where $\Gamma_e^{(0)} = \{e\}$ and $\Gamma_e^{(1)} = \Gamma$. We define the mapping $\tree_e: \Gamma^* \to \T_{\Gamma_e}$ such that $\tree_e(a_1 \cdots a_n) = a_n(a_{n-1}(\ldots a_1(e) \ldots))$; thus $\tree_e(\varepsilon) = e$. Clearly, $\tree_e$ is bijective. Vice versa, given a string ranked alphabet $\Sigma$ with, say, $\Sigma^{(0)} = \{e\}$, then $\tree_e$ has the type $\tree_e: (\Sigma^{(1)})^* \to \T_\Sigma$. Then we have the following transfer result.

\begin{lemma}\rm \label{lm:wsa=wta-over-string-ra} \cite[p.~324]{fulvog09new} (also cf. \cite[Lm.~4.1.3]{fulvog26}) Let $\Gamma$ be an alphabet and $\B$ be a strong bimonoid. Then the following two statements hold.
\begin{compactenum} 
\item[(1)] For every weighted automaton $\cA$ over  $\Gamma$  and  $\B$ and  $e \not\in \Gamma$, we can construct a weighted tree automaton $\cB$ over $\Gamma_e$ and $\B$ such that $\runsem{\cA}=\runsem{\cB}\circ \tree_e$ and $\initialsem{\cA}=\initialsem{\cB}\circ \tree_e$. Thus, in particular, $\tree_e(\supp(\runsem{\cA}) =\supp(\runsem{\cB})$ and $\tree_e(\supp(\initialsem{\cA})) = \supp(\initialsem{\cB})$. 
\item[(2)] For every string ranked alphabet $\Gamma_e$ and weighted tree automaton $\cB$ over $\Gamma_e$ and $\B$, we can construct a weighted automaton $\cA$ over  $\Gamma$  and  $\B$ such that $\runsem{\cA}=\runsem{\cB}\circ \tree_e$ and $\initialsem{\cA}=\initialsem{\cB}\circ \tree_e$.  Thus, in particular, $\supp(\runsem{\cB}) = \tree_e(\supp(\runsem{\cA}))$ and $\supp(\initialsem{\cB}) = \tree_e(\supp(\initialsem{\cA}))$. 
\end{compactenum}
\end{lemma}

Also it is folklore that, essentially, weighted tree automata over the Boolean semiring $\Boole$ are finite-state tree automata (cf., e.g., \cite[Sec.~4.2]{fulvog26}). Here we use this fact for the following definition: a tree language $L \subseteq \T_\Sigma$ is \emph{recognizable} if there exists a weighted tree automaton $\cA$ over $\Sigma$ and $\Boole$ such that $L = \supp(\sem{\cA})$ (for a specification of recognizable tree languages in terms of finite-state tree automata we refer to \cite{gecste84,eng75-15}).

The following result recalls the main known sufficient conditions for a strong bimonoid  $\B$ which ensure that, for each weighted tree automaton  $\cA$ over $\Sigma$ and $\B$, the run support and the initial algebra support of $\cA$ coincide; for the case of positive strong bimonoids cf. \cite[Thm.~4.1]{ghovog23}, also \cite[Lm.~16.2.3]{fulvog26}; the case that  $\B$ is distributive is immediate by Theorem \ref{thm:wta-semiring-runsem=initialsem} because the two semantics' of $\cA$ coincide. 

\begin{theorem} \label{thm:sufficient-cond-wta-equality-support}
  Let $\Sigma$ be a ranked alphabet and let $\B$ be a strong bimonoid. If  $\B$ is positive or distributive, then for each weighted tree automaton $\cA$ over  $\Sigma$  and  $\B$, we have $\supp(\runsem{\cA}) = \supp(\initialsem{\cA})$. 
\end{theorem}

Recall that a non-trivial ranked alphabet is either monadic or branching.
As a straightforward consequence of Lemma~\ref{lm:wsa=wta-over-string-ra} and Theorem \ref{thm:wsa-equivalence}, 
we obtain the characterization of Property~\eqref{eq:equality-of-supports}, given in the Introduction, for weighted tree automata over arbitrary non-trivial monadic ranked alphabets and zero-sum-free strong bimonoids.

\begin{corollary}\label{cor:monadic-ra-wta-support} \rm Let $\Sigma$ be a non-trivial monadic ranked alphabet, and let $\B$  be a zero-sum-free strong bimonoid.
Then the following two statements are equivalent.
\begin{compactenum}
\item[(1)] $\B$  is zero-right-distributive.
\item[(2)] For each weighted tree automaton $\cA$ over $\Sigma$ and $\B$, we have $\supp(\runsem{\cA}) = \supp(\initialsem{\cA})$.
\end{compactenum}
\end{corollary}

\begin{proof}  We have  $\Sigma = \Sigma^{(0)} \cup \Sigma^{(1)}$ and $\Sigma^{(1)}\ne \emptyset$.
For  each weighted tree automaton $\cA$ over $\Sigma$ and $\B$ and for each  $\alpha \in \Sigma^{(0)}$, we consider the string ranked alphabet  $\Sigma_\alpha$ and the restriction $\cA_\alpha$ as in Observation~\ref{obs:restriction-of-wta-to-one-nullary-symbol}.

(1)$\Rightarrow$(2): Let $\cA$ be a weighted tree automaton over $\Sigma$ and $\B$. 
By Lemma~\ref{lm:wsa=wta-over-string-ra} and Theorem \ref{thm:run-=-init}(1)$\Rightarrow$(2), we have  $\supp(\runsem{\cA_\alpha}) = \supp(\initialsem{\cA_\alpha})$ \ for each  $\alpha \in \Sigma^{(0)}$. Hence, by Observation~\ref{obs:restriction-of-wta-to-one-nullary-symbol}(3),
\[\supp(\runsem{\cA}) = \bigcup_{\alpha \in \Sigma^{(0)}} \supp(\runsem{\cA_\alpha})
= \bigcup_{\alpha \in \Sigma^{(0)}} \supp(\initialsem{\cA_\alpha})
= \supp(\initialsem{\cA})\enspace.\]

(2)$\Rightarrow$(1): Choose $\alpha \in \Sigma^{(0)}$. We easily obtain that condition (2) holds 
for each weighted tree automaton  $\cB$ over  $\Sigma_\alpha$  and  $\B$. 
Then Lemma~\ref{lm:wsa=wta-over-string-ra} and Theorem \ref{thm:wsa-equivalence}(2)$\Rightarrow$(1) imply condition (1).
\end{proof}

\subsection{Support results for weighted tree automata on branching ranked alphabets}\label{sec:support-wta-theorems-b}

 The main goal of this section will be the proof of the following tree analogon to Theorem~\ref{thm:wsa-equivalence}. Since monadic ranked alphabets 
have already been considered in Corollary \ref{cor:monadic-ra-wta-support}, here we focus on weighted tree automata over branching ranked alphabets and zero-sum-free strong bimonoids.

        \begin{theorem-rect}  \label{thm:bi-strongly-zsf-equiv-equ-supp-b} Let  $\Sigma$  be a branching ranked alphabet and  $\B$  be a zero-sum-free strong bimonoid.
    Then the following two statements are equivalent.
    \begin{compactenum}
    \item[(1)] $\B$ is zero-distributive.
     \item[(2)] For each weighted tree automaton $\cA$ over $\Sigma$ and $\B$, we have $\supp(\runsem{\cA}) = \supp(\initialsem{\cA})$.
         \end{compactenum}
       \end{theorem-rect}

Note the formal analogy of Theorem~\ref{thm:bi-strongly-zsf-equiv-equ-supp-b} to Theorem~\ref{thm:wta-semiring-runsem=initialsem}: in condition (2), we have weakened the equality of the two semantics to the equality of the two supports, and this corresponds precisely to a weakening of distributivity condition to zero-distributivity. In the same way as in the word case, we can characterize the two inclusions of condition (2) of Theorem~\ref{thm:bi-strongly-zsf-equiv-equ-supp-b}, see Theorems \ref{lm:wta-run-implies-init-b} and \ref{lm:wta-init-implies-run-b}, and the main result is then an immediate consequence of these inclusion characterizations.

In each of the two inclusion characterizations, we employ the weighted tree automaton $\cA$ over the strong bimonoid $\B$   presented in the next example. This is the same wta as the one in the proof of \cite[Lm.~4.1]{rad10}, 1$\Rightarrow$2, (ii) (also cf.   \cite[Thm.~5.3.2]{fulvog22}(B)$\Rightarrow$(A)(ii) and \cite[Lm.~6.3.4]{fulvog26}) except that the root weight of $q$ is $c$ instead of $\1$. For convenience, we recall the wta and its computations where we have replaced $p$ and $q$ by $q_1$ and $q_2$, respectively.

     \begin{definition}\rm \label{ex:A2-b}  Let  $\alpha \in \Sigma^{(0)}$ and $\sigma \in \Sigma^{(k)}$ for some $k \ge 2$.
        Moreover, let  $a,b,b',c \in B$. We can construct a weighted tree automaton $\cA$ over $\Sigma$ and $\B$ such that
\begin{align*}
  \initialsem{\cA}(\sigma(\alpha,\ldots,\alpha,\sigma(\alpha,\ldots,\alpha))) &= a \otimes (b \oplus  b') \otimes c\\
           \runsem{\cA}(\sigma(\alpha,\ldots,\alpha,\sigma(\alpha,\ldots,\alpha))) &= (a \otimes  b \otimes c) \oplus (a \otimes b'\otimes c) \enspace.
           \end{align*}

 We construct $\cA = (Q,\delta,F)$ as follows.   
\begin{compactitem}
\item $Q  = \{q_a, q_b, q_{b'}, q_{\1}, q_1, q_2 \}$,

 \item $\delta_0(\varepsilon,\alpha,q_a) =a$, $\delta_0(\varepsilon,\alpha,q_b) =b$, $\delta_0(\varepsilon,\alpha,q_{b'}) =b'$, $\delta_0(\varepsilon,\alpha,q_{\1}) =\1$, $\delta(\varepsilon,\alpha,q_1)=\delta(\varepsilon,\alpha,q_2)=\0$ and, for every $p_1,\ldots,p_k,p\in Q$, 
\[
\delta_k(p_1\ldots p_k,\sigma,p) = 
\begin{cases}
  \1 &\text{ if } p_1\cdots p_k = q_b\underbrace{q_{\1}\cdots q_{\1}}_{k-1} \text{ and }  p=q_1 \\
  \1 &\text{ if } p_1\cdots p_k = q_{b'}\underbrace{q_{\1}\cdots q_{\1}}_{k-1} \text{ and }  p=q_1,\\
    \1 &\text{ if } p_1\cdots p_k = q_a\underbrace{q_{\1}\cdots q_{\1}}_{k-2}q_1 \text{ and }  p=q_2,\\
\0 &\text{ otherwise.}
\end{cases}
\]
For every $\theta \in \Sigma \setminus \{\alpha, \sigma\}$ with rank $n \in \mathbb{N}$ and every $p_1,\ldots,p_n,p \in Q$, we let $\delta_n(p_1\cdots p_n,\theta,p) = \0$.

\item $F_{q_a}= F_{q_b} = F_{q_{b'}}= F_{\1} =F_{q_1}= \0$ and $F_{q_2} = c$.
\end{compactitem}

\begin{figure}
  \centering

\begin{tikzpicture}[level distance=3.5em,
  every node/.style = {align=center}]]
  \pgfdeclarelayer{bg}    
  \pgfsetlayers{bg,main}  

  \newcommand{\mydista}{1.8mm} 
  \newcommand{\mydistaa}{2.2mm} 
  \newcommand{\mydistb}{0.8mm} 
  \newcommand{\mydistc}{-3.2mm} 
  \newcommand{\mydistd}{-8.9mm} 
  \tikzstyle{mycircle}=[draw, circle, inner sep=-2mm, minimum height=5mm]
  \tikzstyle{mybox}=[draw, ellipse, inner sep=-2mm, minimum height=5mm]

\begin{scope}[level 1/.style={sibling distance=30mm},level 2/.style={sibling distance=25mm}]

 \node at (-0.8, 0.6) {$\xi:$};

 \node (N0) {$\sigma$}
 child {node (N1) {$\alpha$}}
         child {node (N2) {$\alpha$}}
         child {node (N3) {$\sigma$}
           child {node (N4) {$\alpha$}}
           child {node (N5) {$\alpha$}}
         child {node (N6) {$\alpha$}}};

 \node [anchor=west] at ([xshift=\mydista]N0.east) {$\h(\xi)_{q_2} = a \otimes (b \oplus b')$};
 \node [anchor=north] at ([xshift=\mydistc]N1.east) {$\h(\alpha)_{q_a}=a$};
 \node [anchor=north] at ([xshift=\mydistc]N2.east) {$\h(\alpha)_{q_{\1}}=\1$};
 \node [anchor=west] at ([xshift=\mydistb]N3.east) {$\h(\sigma(\alpha,\alpha,\alpha))_{q_1} = b\oplus b'$};
 \node [anchor=north] at ([xshift=\mydistc]N4.east) (bnode) {$\h(\alpha)_{q_b}=b$};
 \node [anchor=north] at ([xshift=\mydistd,yshift=-1.0mm]bnode.east) (bnode) {$\h(\alpha)_{q_{b'}}=b'$};
 \node [anchor=north] at ([xshift=\mydistc]N5.east) {$\h(\alpha)_{q_{\1}}=\1$};
 \node [anchor=north] at ([xshift=\mydistc]N6.east) {$\h(\alpha)_{q_{\1}}=\1$};
\end{scope}
\end{tikzpicture}

\caption{\label{fig:init-2-b} Some values of the form $\h_\cA(\zeta)_p$ where $\cA$ is the weighted automaton of Definition~\ref{ex:A2-b}, $p$ is a state of $\cA$, and $\zeta$ is a subtree of $\xi= \sigma(\alpha,\alpha,\sigma(\alpha,\alpha,\alpha))$, i.e.,  $k=3$.}
\end{figure}   

Let $\xi=\sigma(\xi_1,\ldots,\xi_{k-1},\xi_k)$ where $\xi_i= \alpha$ for each $i \in [k-1]$ and $\xi_k= \sigma(\alpha,\ldots,\alpha)$. Then we can calculate as follows (cf. Figure~\ref{fig:init-2-b}): 
\begingroup
\allowdisplaybreaks
\begin{align*}
 \initialsem{\cA}(\xi) =& \bigoplus_{p \in Q} \h_{\cA}(\xi)_p \otimes F_p=\h_{\cA}(\xi)_{q_2} \otimes c
  \tag{because $F_p = \0$ for each $p \in Q\setminus \{q_2\}$, and $F_{q_2} = c$}\\
  = & \ \Big(\bigoplus_{p_1,\dots,p_k \in Q}  \big(\bigotimes_{i=1}^k \h_{\cA}(\xi_i)_{p_i}\big) \otimes \ \delta_k(p_1\dots p_k,\sigma,q_2)\Big)\otimes c\\
   = & \  \h_{\cA}(\alpha)_{q_a} \otimes \big(\bigotimes_{i=2}^{k-1} \h_{\cA}(\alpha)_{q_{\1}}\big) \otimes \h_{\cA}(\sigma(\alpha,\ldots,\alpha))_{q_1} \otimes \ \delta_k(q_aq_{\1}\dots q_{\1}q_1,\sigma,q_2) \otimes c
       \tag{because  $\delta_k(p_1\cdots p_{k},\sigma,q_2) =\0$ for each $p_1\cdots p_{k} \not= q_aq_{\1}\dots q_{\1}q_1$}\\
  = & \  a \otimes \big(\bigotimes_{i=2}^{k-1} \1 \big)\otimes \h_{\cA}(\sigma(\alpha,\ldots,\alpha))_{q_1} \otimes \ \1 \otimes c \\
 = & \  a \otimes \h_{\cA}(\sigma(\alpha,\ldots,\alpha))_{q_1} \otimes c \\
 = & \  a \otimes \Big(\bigoplus_{p_1,\ldots,p_k \in Q} \bigotimes_{i=1}^k \h_{\cA}(\alpha)_{p_i} \otimes \delta_k(p_1\cdots p_k,\sigma,q_1) \Big) \otimes c \\
  = & \  a \otimes \Big(\ \ \ \h_{\cA}(\alpha)_{q_b} \otimes \Big(\bigotimes_{i=2}^{k} \h_{\cA}(\alpha)_{q_{\1}}\Big) \otimes \delta_k(q_bq_{\1}\cdots q_{\1},\sigma,q_1)  \\
  & \hspace*{7mm} \  \oplus \ \h_{\cA}(\alpha)_{q_{b'}} \otimes \Big(\bigotimes_{i=2}^{k} \h_{\cA}(\alpha)_{q_{\1}}\Big) \otimes \delta_k(q_{b'}q_{\1}\cdots q_{\1},\sigma,q_1) \Big) \otimes c
  \tag{because $\delta_k(p_1\cdots p_{k},\sigma,q_1) = \0$ for each $p_1\cdots p_{k} \not\in\{q_bq_{\1}\cdots q_{\1}, q_{b'}q_{\1}\cdots q_{\1}\}$}\\
 = & \  a \otimes \Big(b \otimes \Big(\bigotimes_{i=2}^{k} \1\Big) \otimes \1  \  \oplus \ b' \otimes \Big(\bigotimes_{i=2}^{k} \1\Big) \otimes \1 \Big) \otimes c\\
  =& \ a \otimes (b \oplus  b') \otimes c \enspace.
\end{align*}
\endgroup

Thus
\begin{equation}\label{equ:wta-specialtree-init-sem-b}
   \initialsem{\cA}(\sigma(\alpha,\ldots,\alpha,\sigma(\alpha,\ldots,\alpha))) = a \otimes (b \oplus  b') \otimes c \enspace.
  \end{equation}

  It is easy to see that $\supp(\initialsem{\cA}) \subseteq \{\sigma(\alpha,\ldots,\alpha,\sigma(\alpha,\ldots,\alpha))\}$. Since $\initialsem{\cA}(\alpha) = \0$, we have  
\begin{equation}\label{equ:wta-im-init-sem-b}
   \im(\initialsem{\cA}) = \{\0, a \otimes (b \oplus  b') \otimes c\} \enspace.
  \end{equation}

We consider the runs $\rho_1,\rho_2 \in \R_{\cA}(\xi)$ (cf. Figure~\ref{fig:run-2-b}) with
\begin{compactitem}
\item $\rho_1(\varepsilon)=q_2$, $\rho_1(1)=q_a$, $\rho_1(i)=q_{\1}$ for every $i$ with $2\leq i\leq k-1$, $\rho_1(k)=q_1$, $\rho_1(k1)=q_b$,\\ $\rho_1(kj)=q_{\1}$ for every $j$ with $2 \le j \le k$, and
\item $\rho_2$ is defined in the same way as $\rho_1$ except that $\rho_2(k1)=q_{b'}$.
\end{compactitem}

\begin{figure}
  \centering

\begin{tikzpicture}[level distance=3.5em,
  every node/.style = {align=center}]]
  \pgfdeclarelayer{bg}    
  \pgfsetlayers{bg,main}  

  \newcommand{\mydista}{1.8mm} 
  \newcommand{\mydistaa}{2.2mm} 
  \newcommand{\mydistb}{0.8mm} 
  \tikzstyle{mycircle}=[draw, circle, inner sep=-2mm, minimum height=5mm]
  \tikzstyle{mycirclebig}=[draw, circle, inner sep=-2mm, minimum height=8mm]
  \tikzstyle{mybox}=[draw, ellipse, inner sep=-2mm, minimum height=5mm]

\begin{scope}[level 1/.style={sibling distance=20mm},level 2/.style={sibling distance=15mm}]

 \node at (-0.8, 0.6) {$\xi:$};
 \node at (1.9, 0.6) {$\rho_1 \in \R_\cA (\xi)$:};

 \node (N0) {$\sigma$}
 child {node (N1) {$\alpha$}}
         child {node (N2) {$\alpha$}}
         child {node (N3) {$\sigma$}
           child {node (N4) {$\alpha$}}
           child {node (N5) {$\alpha$}}
         child {node (N6) {$\alpha$}}};

 \node [mycircle, anchor=west] at ([xshift=\mydista]N0.east) {$q_2$};
 \node [mycircle, anchor=west] at ([xshift=\mydistb]N1.east) {$q_a$};
 \node [mycircle, anchor=west] at ([xshift=\mydistb]N2.east) {$q_{\1}$};
 \node [mycircle, anchor=west] at ([xshift=\mydistb]N3.east) {$q_1$};
 \node [mycircle, anchor=west] at ([xshift=\mydistb]N4.east) {{\small $q_b$}};
 \node [mycircle, anchor=west] at ([xshift=\mydistb]N5.east) {$q_{\1}$};
 \node [mycircle, anchor=west] at ([xshift=\mydistb]N6.east) {$q_{\1}$};
\end{scope}

\begin{scope}[xshift=70mm,level 1/.style={sibling distance=20mm},level 2/.style={sibling distance=15mm}]

 \node at (-0.8, 0.6) {$\xi:$};
 \node at (1.9, 0.6) {$\rho_2 \in \R_\cA (\xi)$:};

 \node (N0) {$\sigma$}
 child {node (N1) {$\alpha$}}
         child {node (N2) {$\alpha$}}
         child {node (N3) {$\sigma$}
           child {node (N4) {$\alpha$}}
           child {node (N5) {$\alpha$}}
         child {node (N6) {$\alpha$}}};

 \node [mycircle, anchor=west] at ([xshift=\mydista]N0.east) {$q_2$};
 \node [mycircle, anchor=west] at ([xshift=\mydistb]N1.east) {$q_a$};
 \node [mycircle, anchor=west] at ([xshift=\mydistb]N2.east) {$q_{\1}$};
 \node [mycircle, anchor=west] at ([xshift=\mydistb]N3.east) {$q_1$};
 \node [mycircle, anchor=west] at ([xshift=\mydistb]N4.east) {{\small $q_{b'}$}};
 \node [mycircle, anchor=west] at ([xshift=\mydistb]N5.east) {$q_{\1}$};
 \node [mycircle, anchor=west] at ([xshift=\mydistb]N6.east) {$q_{\1}$};
\end{scope}
\end{tikzpicture}
  
\caption{\label{fig:run-2-b} Runs $\rho_1$ and $\rho_2$ of $\cA$ on $\sigma(\alpha,\alpha,\sigma(\alpha,\alpha,\alpha))$, cf. Definition~\ref{ex:A2-b} with  $k=3$.}
\end{figure}
Then we have:
\begingroup
\allowdisplaybreaks
\begin{align*}
  & \ \runsem{\cA}(\xi) = \bigoplus_{\rho \in \R_{\cA}(\xi)}\wt_{\cA}(\rho,\xi) \otimes F_{\rho(\varepsilon)}= \big(\wt_{\cA}(\rho_1,\xi) \otimes c\big) \oplus \big(\wt_{\cA}(\rho_2,\xi) \otimes c\big)
    \tag{because $F_p = \0$ for each $p \in Q\setminus \{q_2\}$, and $F_{q_2} = c$} \\[1mm]
  =& \ \ \ \Big(\delta_0(\varepsilon,\alpha,q_a) \otimes \Big(\bigotimes_{i=1}^{k-2}  \delta_0(\varepsilon,\alpha,q_{\1})\Big) \otimes  \delta_0(\varepsilon,\alpha,q_b) \otimes \Big(\bigotimes_{i=1}^{k-1}  \delta_0(\varepsilon,\alpha,q_{\1})\Big)\\
  & \hspace{30mm} \otimes \delta_k(q_bq_{\1} \cdots q_{\1},\sigma,q_1) \otimes \delta_k(q_aq_{\1} \cdots q_{\1}q_1,\sigma,q_2) \otimes c\Big) \\
  & \oplus  \Big(\delta_0(\varepsilon,\alpha,q_a) \otimes \Big(\bigotimes_{i=1}^{k-2}  \delta_0(\varepsilon,\alpha,q_{\1})\Big) \otimes  \delta_0(\varepsilon,\alpha,q_{b'}) \otimes \Big(\bigotimes_{i=1}^{k-1}  \delta_0(\varepsilon,\alpha,q_{\1})\Big)\\
    & \hspace{30mm} \otimes \delta_k(q_{b'}q_{\1} \cdots q_{\1},\sigma,q_1) \otimes \delta_k(q_aq_{\1} \cdots q_{\1}q_1,\sigma,q_2) \otimes c\Big) \\
  =& \ \ \ \Big(a \otimes \Big(\bigotimes_{i=1}^{k-2} \1\Big) \otimes  b  \otimes \Big(\bigotimes_{i=1}^{k-1}  \1)\Big)
     \otimes \1 \otimes \1 \otimes c\Big) \\
  & \oplus  \Big(a \otimes \Big(\bigotimes_{i=1}^{k-2}  \1\Big) \otimes  b' \otimes \Big(\bigotimes_{i=1}^{k-1}  \1\Big)
    \otimes \1 \otimes \1 \otimes c\Big) \\
      =& \ (a \otimes  b \otimes c) \oplus (a \otimes b'\otimes c) \enspace.
\end{align*} 
\endgroup

Thus
\begin{equation}\label{equ:wta-specialtree-run-sem-b}
   \runsem{\cA}(\sigma(\alpha,\ldots,\alpha,\sigma(\alpha,\ldots,\alpha))) = (a \otimes  b \otimes c) \oplus (a \otimes b'\otimes c)\enspace.
  \end{equation}

Also, it  is easy to see that $\supp(\runsem{\cA}) \subseteq \{\sigma(\alpha,\ldots,\alpha,\sigma(\alpha,\ldots,\alpha))\}$. Since $\runsem{\cA}(\alpha)=\0$, we have  
\begin{equation}\label{equ:wta-im-run-sem-b}
   \im(\runsem{\cA}) = \{\0, (a \otimes  b \otimes c) \oplus (a \otimes b'\otimes c)\} \enspace.
  \end{equation}
\hfill $\Box$
        \end{definition}

        The other direction of the inclusion characterizations deduces inclusion relationships between the run support  and the initial algebra support from properties of $\B$. As in the word case, at each position of the given input tree $\xi$, we have to extend the corresponding factor in the weight of a run $\rho_0$ on $\xi$ into a sum (for $\supp(\runsem{\cA}) \subseteq \supp(\initialsem{\cA})$) and reduce the corresponding sum to one summand (for $\supp(\initialsem{\cA}) \subseteq \supp(\runsem{\cA})$), respectively. These transformations are organized by sweeping over the tree $\xi$. Since trees can be branching, the sweeping uses cuts through $\xi$. Each cut is a  sequence of positions of $\xi$ which forms a maximal anti-chain (generalizing single positions in the word case) and it represents a status of the sweep. For the inclusion $\supp(\runsem{\cA}) \subseteq \supp(\initialsem{\cA})$, we start with the cut through all the leaves of $\xi$ (leaves-cut) and proceed towards the cut $(\varepsilon)$, which only consists of the root; for the  inclusion $\supp(\initialsem{\cA}) \subseteq \supp(\runsem{\cA})$, the direction is reversed. 

\begin{figure}[!t]
    \centering
 \begin{tikzpicture}[scale=0.8, every node/.style={transform shape},
					node distance=0.05cm and 0.05cm,
					mycircle/.style={draw, circle, inner sep=0mm, minimum height=5.5mm},
					mydashed/.style={dash pattern=on 2mm off 1mm, thin},
					mydashedarrow/.style={mydashed,->, shorten >=0.1cm,shorten <=0.1cm}],

                                        \begin{scope}[level 1/.style={sibling distance=50mm},
			  level 2/.style={sibling distance=20mm},
			  level 3/.style={sibling distance=18mm},
			  level 4/.style={sibling distance=16mm}]
			  
  \node (n0) at (0,0) {$\eta$}
  child {node (n1) {$\eta$}
    child {node (n11) {$\alpha$}}
      child {node[inner sep=2.5mm] (n12) {$\sigma$}
        child {node (n121) {$\alpha$}}
        child {node (n122) {$\beta$}
        }}
    child {node (n13) {$\alpha$}}}
    child {node[inner sep=2.5mm] (n2) {$\sigma$}
      child {node (n21) {$\sigma$}
        child {node (n121) {$\beta$}}
        child {node (n122) {$\beta$}}}
      child {node (n22) {$\alpha$}} }
    child {node (n3) {$\alpha$}};
  
  \node [mycircle, right = 0.2cm of n0, yshift=0.3cm] (cn0) {$q_0$};
  \node [mycircle, right = 0.1cm of n1] (cn1) {$q_1$};
  \node [mycircle, right = 0.0cm of n11] (cn11) {$q_2$};
  \node [mycircle, right = 0.0cm of n12] (cn12) {$q_3$};
  \node [mycircle, right = 0.0cm of n13] (cn13) {$q_4$};
  \node [mycircle, right = 0.0cm of n2] (cn2) {$q_5$};
  \node [mycircle, right = 0.0cm of n21] (cn21) {$q_6$};
  \node [mycircle, right = 0.0cm of n22] (cn22) {$q_7$};
  \node [mycircle, right = 0.0cm of n3] (cn3) {$q_8$};

  \begin{scope}[every path/.style= mydashed]
    \node [left = 1cm of n11] (leftofn11) {};
    \draw (leftofn11) -- (cn11.west);
    \draw (cn11.east) -- (cn12.west);
    \draw (cn12.east) -- (cn13.west);
    \draw (cn13.east) -- (cn21.west);
    \draw (cn21.east) -- (cn22.west);
    \draw (cn22.east) to[out=0, in=180] (cn3.west) -- (cn3.west);
    \node [right = 1cm of cn3] (rightofcn3) {};
        \draw (cn3.east) -- (rightofcn3)  node[above] {$\kappa$};
  \end{scope}
  
  \node[left= 1.4cm of n0, yshift=0.45cm] {$\xi :$};
  \node[right= 1.4cm of cn0, yshift=0.15cm] {$\rho_0 \in \R_{\cA}(\xi)$};
\end{scope}

\end{tikzpicture}   \caption{\label{fig:1-b} Tree $\xi$ with run $\rho_0$ and  cut $\kappa = (11,12,13,\underline{21,22},3)$.}
\end{figure}

\begin{figure}[!t]
    \centering
 \begin{tikzpicture}[scale=0.8, every node/.style={transform shape},
					node distance=0.05cm and 0.05cm,
					mycircle/.style={draw, circle, inner sep=0mm, minimum height=5.5mm},
					mydashed/.style={dash pattern=on 2mm off 1mm, thin},
					mydashedarrow/.style={mydashed,->, shorten >=0.1cm,shorten <=0.1cm}],

                                        \begin{scope}[level 1/.style={sibling distance=50mm},
			  level 2/.style={sibling distance=20mm},
			  level 3/.style={sibling distance=18mm},
			  level 4/.style={sibling distance=16mm}]
			  
  \node (n0) at (0,0) {$\eta$}
  child {node (n1) {$\eta$}
    child {node (n11) {$\alpha$}}
      child {node[inner sep=2.5mm] (n12) {$\sigma$}
        child {node (n121) {$\alpha$}}
        child {node (n122) {$\beta$}
        }}
    child {node (n13) {$\alpha$}}}
    child {node[inner sep=2.5mm] (n2) {$\sigma$}
      child {node (n21) {$\sigma$}
        child {node (n121) {$\beta$}}
        child {node (n122) {$\beta$}}}
      child {node (n22) {$\alpha$}} }
    child {node (n3) {$\alpha$}};
  
  \node [mycircle, right = 0.2cm of n0, yshift=0.3cm] (cn0) {$q_0$};
  \node [mycircle, right = 0.1cm of n1] (cn1) {$q_1$};
  \node [mycircle, right = 0.0cm of n11] (cn11) {$q_2$};
  \node [mycircle, right = 0.0cm of n12] (cn12) {$q_3$};
  \node [mycircle, right = 0.0cm of n13] (cn13) {$q_4$};
  \node [mycircle, right = 0.0cm of n2] (cn2) {$q_5$};
  \node [mycircle, right = 0.0cm of n3] (cn3) {$q_8$};

  \begin{scope}[every path/.style= mydashed]
    \node [left = 1cm of n11] (leftofn11) {};
    \draw (leftofn11) -- (cn11.west);
    \draw (cn11.east) -- (cn12.west);
    \draw (cn12.east) -- (cn13.west);
    \draw (cn13.east)  to[out=0, in=180] (cn2.west) -- (cn2.west);
    \draw (cn2.east) to[out=0, in=180] (cn3.west) -- (cn3.west);
    \node [right = 1cm of cn3] (rightofcn3) {};
        \draw (cn3.east) -- (rightofcn3)  node[above] {$\kappa'$};
  \end{scope}
  
  \node[left= 1.4cm of n0, yshift=0.45cm] {$\xi :$};
  \node[right= 1.4cm of cn0, yshift=0.15cm] {$\rho_0 \in \R_{\cA}(\xi)$};
\end{scope}

\end{tikzpicture}   \caption{\label{fig:2-b} Tree $\xi$ with run $\rho_0$ and  cut $\kappa' = (11,12,13,\underline{2},3)$.}
\end{figure}

 Figure \ref{fig:1-b} shows a tree $\xi$ and a run $\rho_0$ for which we assume that $\wt(\xi,\rho_0) \otimes F_{\rho_0(\varepsilon)} \not= \0$. Also it exhibits the cut $\kappa = (11,12,13,21,22,3)$, for which we assume that 
  \begin{align*}
    &\h_\cA(\xi|_{11})_{q_2} \otimes \h_\cA(\xi|_{12})_{q_3} \otimes \h_\cA(\xi|_{13})_{q_4} \otimes
    \delta_2(q_2q_3q_4,\eta,q_1)\\
    & \otimes \underbrace{\h_\cA(\xi|_{21})_{q_6} \otimes \h_\cA(\xi|_{22})_{q_7}
    \otimes \delta_2(q_6q_7,\sigma,q_5)}_{\text{subproduct}} \otimes \ \h_\cA(\xi|_3)_{q_8} \otimes \delta_3(q_1q_5q_8,\eta,q_0) \otimes F_{q_0}  \not= \0 \enspace.
  \end{align*}
  That is, in the product
  \[
    \wt_\cA(\xi,\rho_0)= \bigotimes_{\substack{u \in \pos(\xi)\\\text{traversal in $<_\mathrm{po}$}}} \delta_{\rk(\xi(u))}(\rho_0(u1) \cdots \rho_0(u \, \rk(\xi(u))),\xi(u),\rho_0(u))\enspace,
  \]
  for each position $w$ on the cut, the corresponding subproduct is replaced by $\h_\cA(\xi|_w)_{\rho_0(w)}$. 

To prove $\supp(\runsem{\cA}) \subseteq \supp(\initialsem{\cA})$, we proceed to the cut $\kappa' = (11,12,13,\underline{2},3)$ (cf. Figure \ref{fig:2-b}), which results from $\kappa$ by replacing the subsequence $(21,22)$ of all the children of position $2$ by $2$ itself; and we have to prove that 
  \begin{align*}
    &\h_\cA(\xi|_{11})_{q_2} \otimes \h_\cA(\xi|_{12})_{q_3} \otimes \h_\cA(\xi|_{13})_{q_4} \otimes
    \delta_2(q_2q_3q_4,\eta,q_1)\\
    & \otimes \underbrace{\h_\cA(\xi|_{2})_{q_5}}_{\text{sum}} \otimes \ \h_\cA(\xi|_3)_{q_8} \otimes \delta_3(q_1q_5q_8,\eta,q_0) \otimes F_{q_0}  \not= \0 \enspace.
  \end{align*}
This can be achieved as follows. The underbraced subproduct is replaced by the sum 
\[
\bigoplus_{p_1,p_2 \in Q^2} \h_\cA(\xi|_{21})_{p_1} \otimes \h_\cA(\xi|_{22})_{p_2} \otimes \delta_2(p_1p_2,\sigma,q_5) = \h_\cA(\sigma(\sigma(\beta,\beta),\alpha))_{q_5}  = \h_\cA(\xi|_{2})_{q_5} \enspace;
\]
since the subproduct is one of the summands (for $p_1p_2 = q_6q_7$), by assumption on $\B$, the $(\not=\0)$-property is propagated from $\kappa$ to $\kappa'$. This example shows that the condition zero-right-distributive is not sufficient, because there may occur factors to the left of the subproduct; hence we assume that $\B$ is zero-distributive. 
The proof of the other inclusion $\supp(\initialsem{\cA}) \subseteq \supp(\runsem{\cA})$ works in the reversed way. There we have to construct a run on $\xi$, which emerges from the successive choices of summands.

    Formally, let $\xi \in \T_\Sigma$. A \emph{cut through $\xi$} is a sequence $(w_1,\ldots,w_n)$ such that
    \begin{compactitem}
    \item $n \in \mathbb{N}_+$ and $w_i \in \pos(\xi)$, for each $i \in [n]$,
    \item (independent) for every $i,j \in [n]$, we have $w_i \not\in \prefix(w_j)$, and $w_j \not\in \prefix(w_i)$,
    \item (ordered) for every $i,j \in [n]$, if $i<j$, then  $w_i <_{\mathrm{left-of}} w_j$, and
    \item (complete) for each $w \in \pos_{\Sigma^{(0)}}(\xi)$ there exists $i \in [n]$ such that $w_i \in \prefix(w)$.
    \end{compactitem}
    \newcommand{\Cut}{\mathrm{Cut}}
    \newcommand{\lcut}{\mathrm{lcut}}
    We denote by $\Cut(\xi)$ the set of all cuts through $\xi$. Clearly, the set $\Cut(\xi)$ is finite and $(\varepsilon) \in \Cut(\xi)$. The \emph{leaves-cut through $\xi$} is the cut $(w_1,\ldots,w_n)$ uniquely determined by $n = |\pos_{\Sigma^{(0)}}(\xi)|$. We denote the leaves-cut through $\xi$ by $\lcut(\xi)$.

    We define the binary relation $\prec$ on $\Cut(\xi)$ as follows. For every $\kappa_1,\kappa_2 \in \Cut(\xi)$, we let  $\kappa_1 \prec \kappa_2$ if
    \begin{compactitem}
    \item $\kappa_1=(w_1,\ldots,w_n)$ for some $n \in \mathbb{N}_+$ and $w_i \in \pos(\xi)$ and
      \item there exists $i \in [n]$ such that $w_i$ is not a leaf and
    \(\kappa_2 = (w_1,\ldots,w_{i-1},w_i1,\ldots,w_ik,w_{i+1},\ldots,w_n)\)
    where $k = \rk(\xi(w_i))$.
  \end{compactitem}
  Intuitively, $\kappa_2$ is obtained from $\kappa_1$ by replacing a non-leaf position $w_i$ by the sequence of its children. It is clear that $\prec$ is terminating and $\nf_\prec(\Cut(\xi)) = \{\lcut(\xi)\}$. In particular, this implies that it does not matter in which order the positions of $\kappa_1$ are replaced by their children, because each such sequence of expansions will end up in  $\lcut(\xi)$ (i.e., $\prec$ is confluent).

  Also we define the binary relation $\succ$ on $\Cut(\xi)$ to be the inverse of $\prec$, i.e.,  $\succ = \prec^{-1}$.  Intuitively, if $\kappa_1 \succ \kappa_2$, then $\kappa_2$ is obtained from $\kappa_1$ by replacing the sequence $w_i1,\ldots,w_ik$ of $w_i$'s children  by $w_i$. It is clear that $\succ$ is terminating and $\nf_\succ(\Cut(\xi)) = \{(\varepsilon)\}$. Clearly, also $\succ$ is confluent.
    
    \begin{theorem}\rm \label{lm:wta-run-implies-init-b}  Let  $\Sigma$  be a branching ranked alphabet and $\B$ be a zero-sum-free strong bimonoid. Then the following two statements are equivalent.
      \begin{compactenum}
      \item[(1)] $a \otimes b \otimes c \neq \0$ implies $a \otimes (b \oplus b') \otimes c \neq \0$, for all   $a,b,b',c \in B$.
        \item[(2)] For each weighted tree automaton $\cA$ over $\Sigma$ and $\B$, we have $\supp(\runsem{\cA}) \subseteq \supp(\initialsem{\cA})$.
          \end{compactenum}
\end{theorem}

\begin{proof}  (1)$\Rightarrow$(2): Let $\cA$ be a weighted tree automaton over $\Sigma$ and $\B$ and  $\xi \in \T_\Sigma$ such that $\xi \in \supp(\runsem{\cA})$.
  Hence $\bigoplus_{\rho \in \R_\cA(\xi)} \wt(\xi,\rho) \otimes F_{\rho(\varepsilon)} \not= \0$. Thus we can choose a run $\rho_0 \in \R_\cA(\xi)$ such that $\wt(\xi,\rho_0) \otimes F_{\rho_0(\varepsilon)} \not= \0$. 
  
  By Observation \ref{obs:weight-run-explicit},
  \begin{equation}\label{equ:wt-rho0-not-zero-b}
\wt(\xi,\rho_0)  \otimes F_{\rho_0(\varepsilon)} = \big(\bigotimes_{\substack{w \in \pos(\xi):\\\text{traversal in $<_\mathrm{po}$}}} \delta_{\rk(\xi(w))}(\rho_0(w1)\cdots \rho_0(w\,\rk(\xi(w)),\xi(w),\rho_0(w))\Big)  \otimes F_{\rho_0(\varepsilon)}
    \end{equation}

    Let $\alpha$ be an arbitrary symbol in $\Sigma^{(0)}$. For each $\kappa =(w_1,\ldots,w_n)$ in $\Cut(\xi)$, we denote by $\xi[\kappa \leftarrow \alpha]$ the tree which is obtained from $\xi$ by replacing, for each $i \in [n]$, the subtree at position $w_i$ by $\alpha$.  We note that $\{w_1,\ldots,w_n\} \subseteq \pos(\xi[\kappa \leftarrow \alpha])$. Moreover, $\pos(\xi[\kappa \leftarrow \alpha]) \subseteq \pos(\xi)$ and
    \begin{eqnarray}
      &\text{for each $w_1,w_2 \in \pos(\xi[\kappa \leftarrow \alpha]$, we have:}\notag\\
      &\text{$w_1$ occurs left of $w_2$ in the postorder traversal on $\pos(\xi[\kappa \leftarrow \alpha])$ if and only if}\\
      &\text{$w_1$ occurs left of $w_2$ in the postorder traversal on $\pos(\xi)$.\notag}
    \end{eqnarray}

  By well-founded induction on $(\Cut(\xi),\prec)$, we prove the following statement: 
  \begin{eqnarray}\label{equ:wta-h-not-equal-0-b}
    &\text{For every $\kappa \in \Cut(\xi)$, we have:} \
      \Big(\bigotimes_{\substack{w \in \pos(\xi[\kappa \leftarrow \alpha]):\\\text{traversal in $<_\mathrm{po}$}}} b_w\Big)   \otimes F_{\rho_0(\varepsilon)} \not= \0 \enspace,\\
       &\text{where $b_w = \begin{cases}\h_\cA(\xi|_w)_{\rho_0(w)} & \text{if $w \in \{w_1,\ldots,w_n\}$}\\
           \delta_{\rk(\xi(w))}(\rho_0(w1) \cdots \rho_0(w\, \rk(\xi(w))), \xi(w),\rho_0(w)) & \text{otherwise}
           \end{cases}$}\enspace.\notag
\end{eqnarray}

First, let $\kappa = \lcut(\xi)$, i.e., $\kappa$ is the leaves-cut through $\xi$. By definition $\pos(\xi) = \pos(\xi[\lcut(\xi) \leftarrow \alpha])$. Moreover, for each $w \in \pos_{\Sigma^{(0)}}(\xi)$, we have $\h_\cA(\xi|_w)_{\rho_0(w)} = \delta_0(\varepsilon,\xi(w),\rho_0(w))$. Hence, by equation \eqref{equ:wt-rho0-not-zero-b} and since $\wt(\xi,\rho_0) \otimes F_{\rho_0(\varepsilon)} \not= \0$, we obtain
\begin{align*}
  & \Big(\bigotimes_{\substack{w \in \pos(\xi[\lcut(\xi) \leftarrow \alpha]):\\\text{traversal in $<_\mathrm{po}$}}} \!\! b_w\Big) \otimes F_{\rho_0(\varepsilon)}\\
  = &\ \Big(\bigotimes_{\substack{w \in \pos(\xi):\\\text{traversal in $<_\mathrm{po}$}}} \hspace{-6mm} \delta_{\rk(\xi(w))}(\rho_0(w1)\cdots \rho_0(w\,\rk(\xi(w)),\xi(w),\rho_0(w))\Big) \otimes F_{\rho_0(\varepsilon)}\\
  = &\ \wt(\xi,\rho_0) \otimes F_{\rho_0(\varepsilon)} \not=\0 \enspace.
\end{align*}

For the induction step, let $\kappa \in \Cut(\xi)$ such that $\Big(\bigotimes_{\substack{w \in \pos(\xi[\kappa \leftarrow \alpha]):\\\text{traversal in $<_\mathrm{po}$}}} b_w\Big) \otimes F_{\rho_0(\varepsilon)} \not= \0$ (induction hypothesis). Moreover, let $\kappa' \in \Cut(\xi)$ such that $\kappa' \prec \kappa$. Then there exist $w_1,\ldots,w_n \in \pos(\xi)$ and $i \in [n]$ such that
\begin{compactitem}
\item $\kappa = (w_1,\ldots,w_{i-1},\underbrace{w_{i}1,\ldots,w_{i}\, \rk(\xi(w_i))}_{\text{successors of $w_i$}}, w_{i+1},\ldots,w_n)$ and
  \item $\kappa' = (w_1,\ldots,w_{i-1},w_i, w_{i+1},\ldots,w_n)$.
  \end{compactitem}
  We abbreviate $\rk(\xi(w_i))$ by $k$.

 Then there exist $a,c \in B$ such that
  \begin{align*}
    & \Big(\bigotimes_{\substack{w \in \pos(\xi[\kappa \leftarrow \alpha]):\\\text{traversal in $<_\mathrm{po}$}}} b_w\Big) \otimes F_{\rho_0(\varepsilon)}\\
    &= a \otimes \Big(\bigotimes_{j = 1}^{k} b_{w_ij}\Big)
    \otimes \delta_{k}(\rho_0(w_i1) \cdots \rho_0(w_i\, k), \xi(w_i),\rho_0(w_i))
      \otimes c \otimes F_{\rho_0(\varepsilon)}\\
    &= a \otimes \Big(\bigotimes_{j = 1}^{k} \h_\cA(\xi|_{w_j})_{\rho_0(w_j)}\Big)
    \otimes \delta_{k}(\rho_0(w_i1) \cdots \rho_0(w_i\, k), \xi(w_i),\rho_0(w_i))
    \otimes c \otimes F_{\rho_0(\varepsilon)}\enspace. \tag{because each $w_ij$ is on the cut $\kappa$}
  \end{align*}

  By the induction hypothesis, we know that this quantity is not equal to $\0$.  Then we can compute as follows.
  \begingroup
  \allowdisplaybreaks
  \begin{align*}
    &a \otimes \Big(\Big(\bigotimes_{j = 1}^{k} \h_\cA(\xi|_{w_j})_{\rho_0(w_j)}\Big)
      \otimes \delta_{k}(\rho_0(w_i1) \cdots \rho_0(w_i\, k), \xi(w_i),\rho_0(w_i)) \Big)
      \otimes c  \otimes F_{\rho_0(\varepsilon)} \not=\0\\
    \Rightarrow \ \ \  & a \otimes \Big( \bigoplus_{q_1\cdots q_k \in Q^k} \Big(\bigotimes_{j = 1}^{k} \h_\cA(\xi|_{w_j})_{q_j}\Big)
      \otimes \delta_{k}(q_1 \cdots q_k, \xi(w_i),\rho_0(w_i)) \Big)
                         \otimes c  \otimes F_{\rho_0(\varepsilon)} \not=\0 \tag{by Statement (1) of the theorem}\\[2mm]
     \Leftrightarrow \ \ \  & a \otimes \h_\cA(\xi|_{w_i})_{\rho_0(w_i)} \otimes c  \otimes F_{\rho_0(\varepsilon)} \not=\0 \tag{by definition of $\h_\cA(\xi|_{w_i})_{\rho_0(w_i)}$}\\
    \Leftrightarrow \ \ \ & \Big(\bigotimes_{\substack{w \in \pos(\xi[\kappa' \leftarrow \alpha]):\\\text{traversal in $<_\mathrm{po}$}}} b_w\Big)  \otimes F_{\rho_0(\varepsilon)}\not= \0 \enspace.
  \end{align*}
  \endgroup
  This proves equation \eqref{equ:wta-h-not-equal-0-b}. Then we obtain
  \begingroup
  \allowdisplaybreaks
  \begin{align*}
    \0 & \not= \Big(\bigotimes_{\substack{w \in \pos(\xi[(\varepsilon) \leftarrow \alpha]):\\\text{traversal in postorder}}} b_w\Big)  \otimes F_{\rho_0(\varepsilon)} \tag{by choosing $\kappa = (\varepsilon)$ in \eqref{equ:wta-h-not-equal-0-b}} \\
       &= \Big(\bigotimes_{\substack{w \in \pos(\alpha):\\\text{traversal in $<_\mathrm{po}$}}} b_w\Big)  \otimes F_{\rho_0(\varepsilon)}
    \tag{because $\xi[(\varepsilon) \leftarrow \alpha] = \alpha$}\\
      &= b_\varepsilon \otimes F_{\rho_0(\varepsilon)} \tag{because $\pos(\alpha) = \{\varepsilon\}$}\\
    & = \h_\cA(\xi)_{\rho_0(\varepsilon)}  \otimes F_{\rho_0(\varepsilon)} \tag{because $\xi|_\varepsilon = \xi$ and by definition of $b_\varepsilon$}\enspace.  
  \end{align*}
  \endgroup
  Since $\B$ is zero-sum-free, we obtain
  \(\0 \not= \bigoplus_{q\in Q} \h_\cA(\xi)_q \otimes F_q = \initialsem{\cA}(\xi)\).
Hence $\xi \in \supp(\initialsem{\cA}(\xi))$.

\

(2)$\Rightarrow$(1):  Since $\Sigma^{(0)}\not=\emptyset$, we can choose an arbitrary element $\alpha$ from $\Sigma^{(0)}$. Since $\Sigma$ is branching, there exists $k \in \mathbb{N}$ such that $k \ge 2$ and $\Sigma^{(k)}\not= \emptyset$. Let $\sigma$ be a fixed element of $\Sigma^{(k)}$.
We fix $\xi = \sigma(\alpha,\ldots,\alpha,\sigma(\alpha,\ldots,\alpha))$.

Let $a,b,b',c \in B$. We construct the weighted tree automaton $\cA$ over $\Sigma$ and $\B$ as in Definition~\ref{ex:A2-b}. Then we have:
\begin{align*}
  a \otimes b\otimes c \not= \0 \ & \text{ implies } \ (a \otimes b\otimes c) \oplus (a \otimes b'\otimes c) \not= \0 \tag{because $\B$ is zero-sum-free}\\
        & \text{ implies } \runsem{\cA}(\xi)\ne \0 
        \tag{by \eqref{equ:wta-specialtree-run-sem-b}}\\
        & \text{ implies } \initialsem{\cA}(\xi)\ne \0 
        \tag{by Statement (1) of the theorem}\\
        & \text{ implies } \ a \otimes (b\oplus b')\otimes c \not= \0 \enspace. \tag{by \eqref{equ:wta-specialtree-init-sem-b}}
\end{align*}
\end{proof}


The following result characterizes the converse inclusion for the supports.

  \begin{theorem}\rm  \label{lm:wta-init-implies-run-b} Let  $\Sigma$  be a branching ranked alphabet and  $\B$  be a zero-sum-free strong bimonoid.
    Then the following two statements are equivalent.
    \begin{compactenum}
    \item[(1)] $a \otimes (b \oplus b') \otimes c \neq \0$  implies $\big(a \otimes b \otimes c \neq \0$  or 
       $a \otimes b' \otimes c \neq \0\big)$, for all $a,b,b',c \in B$.
     \item[(2)] For each weighted tree automaton $\cA$ over $\Sigma$ and $\B$, we have $\supp(\initialsem{\cA}) \subseteq \supp(\runsem{\cA})$.
         \end{compactenum}
       \end{theorem}

       \begin{proof} (1)$\Rightarrow$(2): Let $\cA$ be a weighted tree automaton  over $\Sigma$ and $\B$ and  $\xi \in \T_\Sigma$ such that $\xi \in \supp(\initialsem{\cA})$. Hence $\bigoplus_{q \in Q} \h_\cA(\xi_q) \otimes F_q \not= \0$. Thus we can choose a state $q_\varepsilon \in Q$ such that $\h_\cA(\xi)_{q_\varepsilon} \otimes F_{q_\varepsilon} \not= \0$. 

  Let $\alpha \in \Sigma^{(0)}$. For each $\kappa =(w_1,\ldots,w_n)$ in $\Cut(\xi)$, we denote by $\xi[\kappa \leftarrow \alpha]$ the tree which is obtained from $\xi$ by replacing, for each $i \in [n]$, the subtree at position $w_i$ by $\alpha$. We note that $\{w_1,\ldots,w_n\} \subseteq \pos(\xi[\kappa \leftarrow \alpha])$.

  By well-founded induction on $(\Cut(\xi),\succ)$, we prove the following statement: 
  \begin{eqnarray}\label{equ:wta-h-not-equal-0-2-b}
    &\text{For every $\kappa \in \Cut(\xi)$, there exists $\rho \in \R_\cA(\xi[\kappa \leftarrow \alpha])$:} \
      \Big(\bigotimes_{\substack{w \in \pos(\xi[\kappa \leftarrow \alpha]):\\\text{traversal in $<_\mathrm{po}$}}} b^\rho_w\Big)   \otimes F_{q_\varepsilon} \not= \0 \enspace,\\
       &\text{where $b_w^\rho = \begin{cases}\h_\cA(\xi|_w)_{\rho(w)} & \text{if $w \in \{w_1,\ldots,w_n\}$}\\
           \delta_{\rk(\xi(w))}(\rho(w1) \cdots \rho(w\, \rk(\xi(w))), \xi(w),\rho(w)) & \text{otherwise}
           \end{cases}$}\enspace.\notag
  \end{eqnarray}
  We note that \eqref{equ:wta-h-not-equal-0-2-b} is similar to \eqref{equ:wta-h-not-equal-0-b} except that now we have to construct a run $\rho$; in \eqref{equ:wta-h-not-equal-0-b} the run $\rho_0$ is given from the beginning. 

First, let $\kappa = (\varepsilon)$. By definition $\pos(\xi[(\varepsilon) \leftarrow \alpha]) = \pos(\alpha) = \{\varepsilon\}$. We define $\rho \in \R_\cA(\xi[(\varepsilon) \leftarrow \alpha])$ such that $\rho(\varepsilon) = q_\varepsilon$ (where $q_\varepsilon$ is specified above). Then  
\begin{align*}
  \Big(\bigotimes_{\substack{w \in \pos(\xi[(\varepsilon) \leftarrow \alpha]):\\\text{traversal in $<_\mathrm{po}$}}} \!\! b^\rho_w\Big) \otimes F_{q_\varepsilon} = b^\rho_\varepsilon \otimes F_{q_\varepsilon} = \h_\cA(\xi|_\varepsilon)_{\rho(\varepsilon)} \otimes F_{q_\varepsilon} = \h_\cA(\xi)_{q_\varepsilon} \otimes F_{q_\varepsilon} \enspace.
\end{align*}
By the above, this quantity is not equal to $\0$.

\

For the induction step, let $\kappa \in \Cut(\xi)$ and $\rho \in \R_\cA(\xi[\kappa \leftarrow \alpha])$ such that $\Big(\bigotimes_{\substack{w \in \pos(\xi[\kappa \leftarrow \alpha]):\\\text{traversal in $<_\mathrm{po}$}}} b^\rho_w\Big) \otimes F_{q_\varepsilon} \not= \0$ (induction hypothesis). Moreover, let $\kappa' \in \Cut(\xi)$ such that $\kappa' \succ \kappa$. Then there exist $w_1,\ldots,w_n \in \pos(\xi)$ and $i \in [n]$ such that
\begin{compactitem}
\item $\kappa' = (w_1,\ldots,w_{i-1},\underbrace{w_{i}1,\ldots,w_{i}\, \rk(\xi(w_i))}_{\text{successors of $w_i$}}, w_{i+1},\ldots,w_n)$ and
  \item $\kappa = (w_1,\ldots,w_{i-1},w_i, w_{i+1},\ldots,w_n)$.
  \end{compactitem}
  We abbreviate $\rk(\xi(w_i))$ by $k$.

    Then there exist $a,c \in B$ such that
  \begin{align*}
    \Big(\bigotimes_{\substack{w \in \pos(\xi[\kappa \leftarrow \alpha]):\\\text{traversal in $<_\mathrm{po}$}}} b^\rho_w\Big) \otimes F_{q_\varepsilon}
    = a \otimes \h_\cA(\xi|_{w_i})_{\rho(w_i)} \otimes c \otimes F_{q_\varepsilon}
  \end{align*}

          By our induction hypothesis, we know that this quantity is not equal to $\0$.  Then we can compute as follows.
  \begingroup  
  \allowdisplaybreaks
  \begin{align*}
    & a \otimes \h_\cA(\xi|_{w_i})_{\rho(w_i)} \otimes c  \otimes F_{q_\varepsilon} \not=\0 \\
\Leftrightarrow \ \ \     &a \otimes \Big( \bigoplus_{q_1\cdots q_k \in Q^k}\Big(\bigotimes_{j = 1}^{k} \h_\cA(\xi|_{w_ij})_{q_j}\Big)
      \otimes \delta_{k}(q_1 \cdots q_k, \xi(w_i),\rho(w_i)) \Big)
                            \otimes c  \otimes F_{q_\varepsilon} \not=\0
    \tag{by definition of $\h_\cA(\xi|_{w_i})_{\rho(w_i)}$}\\
    \Rightarrow \ \ \  & a \otimes \Big(\bigotimes_{j = 1}^{k} \h_\cA(\xi|_{w_ij})_{q_{w_ij}}\Big)
      \otimes \delta_{k}(q_{w_i1} \cdots q_{w_ik}, \xi(w_i),\rho(w_i))
                         \otimes c  \otimes F_{q_\varepsilon} \not=\0
                         \tag{by Statement (1) of the theorem such $q_{w_i1},\ldots, q_{w_ik}$ exist}\\
     \Leftrightarrow \ \ \  & a \otimes \Big(\bigotimes_{j = 1}^{k} b^{\rho'}_{w_ij}\Big)
                              \otimes \delta_{k}(q_{w_i1} \cdots q_{w_ik}, \xi(w_i),\rho(w_i)) \otimes c  \otimes F_{q_\varepsilon} \not=\0
                              \tag{where $\rho'|_{\pos(\xi[\kappa \leftarrow \alpha])} = \rho$ and $\rho'(w_ij) = q_{w_ij}$ for each $j \in [k]$ }\\
    \Leftrightarrow \ \ \ & \Big(\bigotimes_{\substack{w \in \pos(\xi[\kappa' \leftarrow \alpha]):\\\text{traversal in $<_\mathrm{po}$}}} b^{\rho'}_w\Big)  \otimes F_{q_\varepsilon}\not= \0 \enspace.
  \end{align*}
  \endgroup
  This proves equation \eqref{equ:wta-h-not-equal-0-2-b}. Then we obtain
  \begingroup
  \allowdisplaybreaks
  \begin{align*}
    \0 &\not= \Big(\bigotimes_{\substack{w \in \pos(\xi[\lcut(\xi) \leftarrow \alpha]):\\\text{traversal in $<_\mathrm{po}$}}} b^\rho_w\Big)   \otimes F_{q_\varepsilon} \tag{choosing $\kappa = \lcut(\xi)$; note that $\rho \in \R_\cA(\xi)$}\\
       &= \Big(\bigotimes_{\substack{w \in \pos(\xi):\\\text{traversal in $<_\mathrm{po}$}}} b^\rho_w\Big)   \otimes F_{q_\varepsilon}\\
       &= \Big(\bigotimes_{\substack{w \in \pos(\xi):\\\text{traversal in $<_\mathrm{po}$}}} \delta_{\rk(\xi(w))}(\rho(w1) \cdots \rho(w\, \rk(\xi(w))), \xi(w),\rho(w)) \Big)   \otimes F_{q_\varepsilon}\\
    &= \wt(\rho,\xi)  \otimes F_{q_\varepsilon} \enspace.
    \end{align*}
  \endgroup

  Since $\B$ is zero-sum-free, we also have
  \(\runsem{\cA}(\xi) = \bigoplus_{\rho \in \R_\cA(\xi)} \wt(\rho,\xi) \otimes F_{\rho(\varepsilon)} \not= \0 \).
  Hence $\xi \in \supp(\runsem{\cA})$.
  
\

\underline{(2)$\Rightarrow$(1):} Since $\Sigma^{(0)}\not=\emptyset$, we can choose an arbitrary element from $\Sigma^{(0)}$; let us call it $\alpha$. Since $\Sigma$ is branching, there exists $k \in \mathbb{N}$ such that $k \ge 2$ and $\Sigma^{(k)}\not= \emptyset$. Let $\sigma$ be a fixed element of $\Sigma^{(k)}$.
We fix $\xi = \sigma(\alpha,\ldots,\alpha,\sigma(\alpha,\ldots,\alpha))$.

Let $a,b,b',c \in B$. We construct the weighted tree automaton $\cA$ over $\Sigma$ and $\B$ as in Definition~\ref{ex:A2-b}. Then   we obtain:
\begin{align*}
a \otimes  (b\oplus b')\otimes c \not= \0 
& \text{ implies } \ \initialsem{\cA}(\xi) \ne \0 
\tag{by \eqref{equ:wta-specialtree-init-sem-b}}\\
& \text{ implies } \ \runsem{\cA}(\xi) \ne \0 
\tag{by Statement (1) of the theorem}\\
\ & \text{ implies } \ (a \otimes b \otimes c) \oplus (a \otimes b'\otimes c)\not=\0 
\tag{by \eqref{equ:wta-specialtree-run-sem-b}}\\[2mm]
                                 & \text{ implies } \ a \otimes b \otimes c \neq \0 \text{ or } a \otimes b' \otimes c \neq \0 \enspace. \qedhere
\end{align*}
 \end{proof}

We obtain the following equivalence for commutative zero-sum-free strong bimonoids, which is an immediate consequence of Observation  \ref{obs:com-zero-right-distr-implies-zero-distr} and Theorems \ref{thm:wsa-equivalence} and \ref{thm:bi-strongly-zsf-equiv-equ-supp-b}.

 \begin{corollary}\rm \label{cor:comm-zsf-wsa=wta-b} Let $\B$ be a zero-sum-free commutative or a zero-sum-free left-distributive strong bimonoid, and let $\Gamma$ be an alphabet and $\Sigma$ a branching ranked alphabet. Then the following statements are equivalent.
  \begin{compactenum}
    \item[(1)] For every weighted automaton $\cA$ over  $\Gamma$  and  $\B$, we have $\supp(\runsem{\cA}) = \supp(\initialsem{\cA})$.
  \item[(2)] For every weighted tree automaton $\cA$ over $\Sigma$ and $\B$, we have $\supp(\runsem{\cA}) = \supp(\initialsem{\cA})$.
    \end{compactenum}
  \end{corollary}

  Next, we can use the tree analogon of \cite[Thm.~3.1]{kir11} (also cf. \cite[Thm.~1]{kir09}) for proving that   initial algebra support languages of weighted tree automata  over branching ranked alphabets and zero-sum-free and zero-distributive commutative strong bimonoids are recognizable tree languages.
  
 \begin{corollary}\rm Let $\B$ be a zero-sum-free zero-distributive commutative strong bimonoid, $\Sigma$ a branching ranked alphabet and $\cA$ a weighted tree automaton over $\Sigma$ and $\B$. Then the tree language $\supp(\initialsem{\cA})$ is recognizable.
 \end{corollary}

 \begin{proof} By Theorem \ref{thm:bi-strongly-zsf-equiv-equ-supp-b}, we have $\supp(\initialsem{\cA}) = \supp(\runsem{\cA})$.
 Now, since  $\B$ is zero-sum-free and commutative, by \cite[Thm.~4.4]{fulhervog18} (\cite[Thm.~16.2.10]{fulvog26}, also cf. \cite[Thm.~4.1]{droheu15} and \cite[Thm.~4.7]{goe17} for the case of unranked trees),  $\supp(\runsem{\cA})$ is  recognizable. 
   \end{proof}

 A strong bimonoid  $\B = (B,\oplus,\otimes,\0,\1)$ is said to be \emph{computable}, if there are effectively given representations of its elements and operations and there is an effective test whether two representations give the same element. Let  $\B$ be computable. Given an alphabet $\Gamma$, a weighted automaton $\cA$ over  $\Gamma$  and  $\B$  and some  $w \in \Gamma^*$, one might want to decide whether  $\runsem{\cA}(w) \neq \0$; similarly, given a ranked alphabet $\Sigma$, a weighted tree automaton $\cA$ over $\Sigma$ and $\B$, and some  $\xi \in \T_\Sigma$, one might want to decide whether $\runsem{\cA}(\xi) \neq \0$. Clearly, this is decidable: we may compute  $\runsem{\cA}(w)$  respectively  $\runsem{\cA}(\xi)$ and then test whether this value is $\0$. But, using the definition of the run semantics, this computation might require exponentially many (in terms of the length of  $w$ resp. the size of $\xi$) multiplications. Now, for zero-sum-free strong bimonoids which are zero-right-distributive resp. zeo-distributive, we can use the initial algebra semantics to achieve a much more efficient decision algorithm. 

 \begin{corollary}\rm \label{cor:computable-zsf-linear-support} Let $\B$ be a computable zero-sum free strong bimonoid. 
  \begin{compactenum}
    \item[(1)] Let  $\B$ be zero-right-distributive. Given an alphabet $\Gamma$, a weighted automaton $\cA$ over  $\Gamma$  and  $\B$ and some $w \in \Gamma^*$, it can be decided with a number of additions or multiplications of $\B$ which is linear in the length of $w$ whether $\runsem{\cA}(w) \neq \0$.
  \item[(2)] Let  $\B$ be zero-distributive. Given a branching ranked alphabet $\Sigma$, a weighted tree automaton $\cA$ over  $\Sigma$  and  $\B$ and some $\xi \in \T_\Sigma$, it can be decided with a number of additions or multiplications of $\B$ which is linear in the size of $\xi$ whether $\runsem{\cA}(\xi) \neq \0$.
    \end{compactenum}
  \end{corollary}
  
  \begin{proof} Let $\cA$ have state set $Q$. In both cases, Theorems \ref{thm:wsa-equivalence} and \ref{thm:bi-strongly-zsf-equiv-equ-supp-b} give $\supp(\runsem{\cA}) = \supp(\initialsem{\cA})$. We compute $\initialsem{\cA}(w)$  respectively $\initialsem{\cA}(\xi)$. In case of (1), this requires $n$ multiplications of a vector in  $B^Q$  with a matrix from $B^{Q\times Q}$, where $n$ is the length of $w$. Hence the total number of operations from $B$ needed to obtain the value of $\initialsem{\cA}(w)$ is linear in $n$.
  In case of~(2), by a similar reasoning, see \cite[Thm.~6.1.1]{fulvog26} and the explicit algorithm given there, the value of $\initialsem{\cA}(\xi)$ can be obtained with a number of operations from $\B$ which is linear in the size of $\xi$. 
  
  Finally, decide whether these values are equal to $\0$. This decides whether $w \in \supp(\runsem{\cA})$ respectively whether $\xi \in \supp(\runsem{\cA})$.
  \end{proof}

\section{Images of initial algebra semantics and run semantics}

In this section, we show that we can extend both Theorem~\ref{thm:wsa-semiring-run=initial} and Theorem \ref{thm:wta-semiring-runsem=initialsem} by a further equivalent condition. This exploits Definition~\ref{ex:special-wsa} respectively Definition~\ref{ex:A2-b} and, in particular,  the equality \eqref{equ:wsa-im-sem} (for the word case)  and the equalities \eqref{equ:wta-im-init-sem-b} and \eqref{equ:wta-im-run-sem-b} (for the tree case).

First, for the word case, we extend Theorem~\ref{thm:wsa-semiring-run=initial} as follows.

\begin{theorem-rect} \label{thm:wsa-im-equ-implies-equ} Let  $\B$ be a strong bimonoid. The following three statements are equivalent:
\begin{compactenum}
\item[(1)] $\B$ is right-distributive.
\item[(2)]  For every alphabet $\Gamma$ and weighted automaton $\cA$ over  $\Gamma$  and  $\B$, we have $\im(\runsem{\cA}) = \im(\initialsem{\cA})$.
\item[(3)] For every alphabet $\Gamma$ and weighted automaton $\cA$ over  $\Gamma$  and  $\B$, we have $\runsem{\cA} = \initialsem{\cA}$.
\end{compactenum}
\end{theorem-rect}
   \begin{proof} (1)$\Rightarrow$(3): Immediate by Theorem~\ref{thm:wsa-semiring-run=initial}(1)$\Rightarrow$(2).
       
         (3)$\Rightarrow$(2): This is trivial.

         (2)$\Rightarrow$(1): Let $a,b,c \in B$ be arbitrary. Consider the corresponding weighted automaton $\cA$ from Example~\ref{ex:special-wsa}. By \eqref{equ:wsa-im-sem}, we have 
\[\im(\runsem{\cA}) = \{\0, a \otimes c \oplus b \otimes c\} \ \ \text{and} \ \ \im(\initialsem{\cA}) =  \{\0, (a \oplus b) \otimes c\}\]
By assumption, $\im(\runsem{\cA}) = \im(\initialsem{\cA})$. In particular, these sets are both singleton sets or have both two elements. In both cases, we obtain $a \otimes c \oplus b \otimes c = (a \oplus b) \otimes c$, and $\B$ is right-distributive.
\end{proof}

Secondly, for the tree case, we can easily lift Theorem \ref{thm:wsa-im-equ-implies-equ}(1)$\Leftrightarrow$(2) from string ranked alphabets to ranked alphabets which are non-trivial and monadic.

\begin{corollary}\label{cor:string-ra-wta-im} \rm Let $\Sigma$ be a non-trivial monadic ranked alphabet, and let $\B$  be a strong bimonoid. 
Then the following two statements are equivalent.
    \begin{compactenum}
    \item[(1)] $\B$ is right-distributive.
    \item[(2)] For each weighted tree automaton $\cA$ over $\Sigma$ and $\B$, we have $\im(\runsem{\cA}) = \im(\initialsem{\cA})$.
         \end{compactenum}
       \end{corollary} 
       
       \begin{proof} We have  $\Sigma = \Sigma^{(0)} \cup \Sigma^{(1)}$ and $\Sigma^{(1)}\not=\emptyset$.
For each $\alpha \in \Sigma^{(0)}$ and each weighted tree automaton $\cA$ over $\Sigma$ and $\B$, we define $\Sigma_\alpha$ and $\cA_\alpha$ as in Observation~\ref{obs:restriction-of-wta-to-one-nullary-symbol}.

(1)$\Rightarrow$(2): Let $\cA$ be a weighted tree automaton over $\Sigma$ and $\B$. By Lemma~\ref{lm:wsa=wta-over-string-ra} and Theorem~\ref{thm:wsa-im-equ-implies-equ}(1)$\Rightarrow$(2), we have  $\im(\runsem{\cA_\alpha}) = \im(\initialsem{\cA_\alpha})$ \ for each  $\alpha \in \Sigma^{(0)}$. Hence, by Observation~\ref{obs:restriction-of-wta-to-one-nullary-symbol}(4),
\[\im(\runsem{\cA}) = \bigcup_{\alpha \in \Sigma^{(0)}} \im(\runsem{\cA_\alpha})
= \bigcup_{\alpha \in \Sigma^{(0)}} \im(\initialsem{\cA_\alpha})
= \im(\initialsem{\cA})\enspace.\]

(2)$\Rightarrow$(1): Choose $\alpha \in \Sigma^{(0)}$. We easily obtain that condition (2) holds 
for each weighted tree automaton  $\cB$ over  $\Sigma_\alpha$  and  $\B$. 
Then Lemma~\ref{lm:wsa=wta-over-string-ra} and  Theorem~\ref{thm:wsa-im-equ-implies-equ}(2)$\Rightarrow$(1) imply condition (1).
\end{proof}

Thirdly, for the case of branching tree alphabets, we extend Theorem \ref{thm:wta-semiring-runsem=initialsem} as follows.

\begin{theorem-rect}\label{thm:wta-im-equ-implies-equ}
 Let  $\Sigma$  be a branching ranked alphabet and $\B$ a strong bimonoid.
    Then the following three statements are equivalent.
    \begin{compactenum}
    \item[(1)]  $\B$ is distributive.
     \item[(2)] For each weighted tree automaton $\cA$ over $\Sigma$ and $\B$, we have $\im(\runsem{\cA}) = \im(\initialsem{\cA})$.
       \item[(3)] For each weighted tree automaton $\cA$ over $\Sigma$ and $\B$, we have $\runsem{\cA} = \initialsem{\cA}$.
         \end{compactenum}
       \end{theorem-rect}

       \begin{proof} (1)$\Rightarrow$(3): Immediate by Theorem \ref{thm:wta-semiring-runsem=initialsem}(1)$\Rightarrow$(2).
       
         (3)$\Rightarrow$(2): This is trivial.

         (2)$\Rightarrow$(1): 
Let $a,b,b',c \in B$. We construct the weighted tree automaton $\cA$ over $\Sigma$ and $\B$ as in Definition~\ref{ex:A2-b}.
Then we have
\begin{align*}
  \im(\initialsem{\cA}) &= \{\0, a \otimes (b \oplus  b') \otimes c \} 
  \tag{by \eqref{equ:wta-im-init-sem-b}}\\
    \im(\runsem{\cA}) &= \{\0, (a \otimes  b \otimes c) \oplus (a \otimes b' \otimes c)\} \enspace. \tag{by \eqref{equ:wta-im-run-sem-b}}
           \end{align*}
By assumption (2), we obtain 
$a \otimes (b \oplus  b') \otimes c = (a \otimes  b \otimes c) \oplus (a \otimes b' \otimes c)$, 
hence $\B$ is distributive.
\end{proof}

\bibliographystyle{plain}


\end{document}